\documentclass[]{jfm}

\usepackage{graphicx}
\usepackage{newtxtext}
\usepackage{newtxmath}
\usepackage{natbib}
\usepackage{color}
\usepackage{hyperref}
\hypersetup{
    colorlinks = true,
    urlcolor   = blue,
    citecolor  = black,
}

\newcommand{\RomanNumeralCaps}[1]
\linenumbers

\title{Multi-scale invariant solutions in plane Couette flow: a reduced-order model approach}
\author{Matthew McCormack\aff{1}\textsuperscript{,}\aff{2}\textsuperscript{,}\corresp{\email{matthew.mccormack@ed.ac.uk}},
  Andr\'e V. G. Cavalieri\aff{3}
\and Yongyun Hwang\aff{2}}

\affiliation{\aff{1}School of Mathematics and Maxwell Institute for Mathematical Sciences, University of Edinburgh, Edinburgh EH9 3FD, UK\aff{2}Department of Aeronautics, Imperial College London, London SW7 2AZ, UK
\aff{3}Divisão de Engenharia Aeroespacial, Instituto Tecnológico de Aeronáutica, São José dos Campos, SP
12228-900, Brazil}

\begin{document}
\maketitle

\begin{abstract}
Plane Couette flow at $\Rey=1200$ (based on the channel half-height and half the velocity difference between the top and bottom plates) is investigated {with a spatial domain designed to retain only two spanwise integral length scales. In this system,} the computation of invariant solutions that are physically representative of the turbulent state has been understood to be challenging. To address this challenge, our approach is to employ an accurate reduced-order model with 600 degrees of freedom (Cavalieri \& Nogueira, \textit{Phys. Rev. Fluids}, vol. 7, 2022, L102601). Using the two-scale energy budget and the temporal cross-correlation of key observables, it is first demonstrated that the model contains most of the multi-scale physical processes identified recently (Doohan \textit{et al.}, \textit{J. Fluid Mech.}, vol. 913, 2021, A8): i.e. the large- and small-scale self-sustaining processes, the energy cascade for turbulent dissipation, and an energy-cascade mediated small-scale production mechanism. Invariant solutions of the reduced-order model are subsequently computed, including 96 equilibria and 43 periodic orbits. It is found that {none of the} computed equilibrium solutions are able to reproduce {an accurate} energy balance associated with the multi-scale dynamics of turbulent state. Incorporation of unsteadiness into invariant solutions is seen to be essential for a sensible description of the multi-scale turbulent dynamics and the related energetics, at least in this type of flow, as periodic orbits with a sufficiently long period are mainly able to describe the complex spatiotemporal dynamics associated with the known multi-scale phenomena.
\end{abstract}



\section{Introduction} \label{sec:introduction}
The theory of chaotic dynamical systems has played an increasing role in the study of turbulence in recent years, stemming from the work of \citet{hopf1948mathematical} who envisaged the state space of a fluid flow as that of the infinite-dimensional function space of the Navier-Stokes equations.  In this view, each unique flow field is represented by a single point in the state space, with a time evolving flow tracing a continuous curve.  As in traditional finite-dimensional dynamical systems theory, this trajectory through the state space is largely influenced and organised by invariant solutions such as equilibria, periodic orbits, and higher dimensional tori, which attract and repel trajectories along their stable and unstable manifolds.  

This approach has seen notable success in describing the laminar-turbulent transition in parallel wall-bounded shear flows that exhibit a sub-critical transition \cite[e.g.][]{Kerswell2005,Eckhardt2007,kawahara2012significance,graham2021exact}, especially when the computational domain of interest is sufficiently small so that the concept of `space' in transition can be ignored {-- note that when the domain size is very large, the spatial evolution of disturbances becomes important and transition has been understood to be better described by the framework of statistical mechanics, such as directed percolation \cite[for this issue, see][]{Barkley2016}.} In this case, the transition, induced by finite amplitude perturbations, can be viewed to be initiated by a saddle-node bifurcation, through which a pair of non-trivial exact solutions to the Navier-Stokes equations emerge. As Reynolds number is increased, the solution with higher skin friction (upper branch) subsequently experiences a bifurcation cascade involving homoclinic tangency \cite[e.g.][]{Kreilos_Eckhardt2012,Lustro2019} and evolves into a chaotic saddle (i.e. a chaotic state with finite lifetime). On the contrary, the solution with lower skin friction (lower branch) often becomes part of the laminar-turbulent separatrix, referred to as \emph{edge of turbulence}, a high-dimensional surface in the state space, where a relative attractor, known as the \emph{edge state}, lives \cite[]{itano2001dynamics,Skufca.96.174101,schneider2007turbulence}. 

The application of dynamical systems concepts to wall-bounded shear flows has been largely restricted to the minimal flow unit of \citet{jimenez1991minimal}. This flow unit, periodic in both the streamwise and spanwise directions, represents the smallest near-wall flow domain that can sustain turbulence, in units based on the viscous inner length scale ($\delta_\nu = \nu/u_\tau$, where $\nu$ is the kinematic viscosity and $u_\tau$ the friction velocity), hereafter referred to with a $\cdot^+$ superscript.  This box size is directly related to the spanwise spacing of streaks ($\lambda^+_z \approx 100$) found in the near-wall region of turbulent flows, first identified experimentally by \citet{kline1967structure}.  Therefore, by construction, the minimal flow unit isolates a single pair of low- and high-speed streaks and captures the related minimal dynamics required to sustain turbulence, often referred to as the self-sustaining process \citep{hamilton_kim_waleffe_1995}.  

This represents a quasi-cyclic process whereby streaks are first generated by streamwise vortices that take energy from the mean shear through the lift-up effect \citep{moffatt1965interaction, ellingsen1975stability,landahl1980note, landahl1990sublayer, butler1993optimal,del2006linear,cossu2009optimal, pujals2009note,hwang2010linear,mckeon2010critical}.  The streaks then undergo a normal-mode instability and/or transient growth, forming downstream undulation of streaks, which subsequently promote the generation of streamwise vortices as a result of nonlinear self-interactions \citep{hamilton_kim_waleffe_1995,schoppa2002coherent,park2011stability,hwang_bengana_2016, cassinelli2017streak,lozano2021cause}.

From a dynamical systems perspective, the self-sustaining process has firmly been associated with a variety of non-trivial invariant solutions that have been computed in plane Couette flow \citep{nagata_1990,clever_busse_1997,kawahara_kida_2001,viswanath2007recurrent,gibson2009equilibrium,cvitanovic2010geometry, hall2010streamwise,deguchi2013emergence}, as well as in other canonical shear flows such as pipe flow \citep{faisst2003traveling,wedin2004exact,willis2016symmetry}, plane Poiseuille flow \citep{waleffe1998three,waleffe_2001,park2015exact,hwang2016invariant,yang2019exact}, stress-driven shear flows \citep{doohan_willis_hwang_2019,doohan_2022-tw} and boundary layer flow \citep{khapko2013localized, kreilos2013edge, deguchi2014free}. In the minimal flow unit, these invariant solutions have been shown to form a skeleton for the chaotic dynamics of the flows \citep{gibson2008visualizing,willis2016symmetry,doohan_willis_hwang_2019}, and are typically seen to capture the underlying mathematical structure and dynamics of the coherent structures. These solutions typically have both stable and unstable manifolds which attract and repel chaotic trajectories throughout the state space, with turbulence being viewed as a chaotic walk between these solutions \citep{kawahara2012significance, graham2021exact}. In this sense, periodic orbits have been understood to provide a low-dimensional representation of a smooth dynamical system, and suitable averaging over these solutions has the ability to yield a meaningful statistical description of the chaotic state \citep{artuso1990recycling, chandler2013invariant,yalniz2021coarse,page2022recurrent}.

As the Reynolds number is increased, it has been shown that the self-sustaining process possibly exists at each of the integral length scales varying from the viscous inner to the large-scale outer units in wall-bounded turbulence \citep{hwang2010self,flores2010hierarchy,hwang2011self,hwang_2015,hwang_bengana_2016}. At each integral length scale, the self-sustaining process supports energy-containing coherent structures, composed of streaks and streamwise vortices, and they are statistically and dynamically self-similar with respect to the integral length scale, consistent with the attached eddy hypothesis of \citet{Townsend1956}. There have been efforts to characterise the possible interactions between these self-similar coherent structures and the related turbulent energy transfer \citep{cho_hwang_choi_2018,lee_moser_2019}, but the dynamical processes underpinning these scale interactions currently remain poorly understood. 

The simplest system containing the dynamics of these interactions was recently studied by \citet{doohan_willis_hwang_2021}, who considered a computational domain twice greater than the minimal flow unit, referred to as the minimal multi-scale flow unit. By construction, this domain only admits the coherent structures at two integral length scales (i.e. $\lambda_z^+\simeq 100, 200$). Several important multi-scale processes have been numerically identified: 1) large- and small-scale self-sustaining processes; 2) energy cascade via the streak instability (or transient growth) and breakdown; 3) energy transport from the large to small scales which drives small-scale turbulent production; 4) the feeding of energy from small to large scales \citep{doohan_willis_hwang_2021}. Furthermore, a number of equilibria and two periodic orbits in this system were computed \citep{doohan_2022-tw}. However, none of them were representative of the full system's chaotic dynamics, as had been seen previously in the minimal flow unit -- most of them were found to contain the single scale dynamics only associated with the corresponding self-sustaining process, and they did not appear to be key players forming the state space structure of turbulence, since they lacked some of the other physically relevant processes. 

The observations of \cite{doohan_2022-tw} {raise} some questions regarding the structure of the state space of turbulence at high Reynolds numbers. It has often been believed that {the closure of a set of relevant} invariant solutions of the Navier-Stokes equations form the ergodic subspace of the turbulent state, in the sense that the solution trajectory visits or shadows many of these solutions at least for some amount of time. This is an important requirement for a meaningful statistical description of the turbulent state in terms of the invariant solutions. However, as the Reynolds number is increased, many of these solutions, especially those which are relatively easily accessible through the numerical continuation of their counterpart computed at lower Reynolds numbers, are no longer directly relevant to turbulence, as the solution trajectory appears to very rarely visit them. This now emphasises the necessity of capturing the invariant solutions buried around the turbulent state at high Reynolds numbers. Such solutions are expected to contain the full multi-scale features of the turbulent state like the one recently obtained by \citet{motoki2021multi} for thermal convection. 

Computation of multi-scale invariant solutions organised around the turbulent state, however, appears to be increasingly challenging as the Reynolds number is increased. This is directly connected to the extremely high computational expense required to find these solutions, as the computation typically involves a large number of iterative direct numerical simulations (DNS) combined with a suitable searching algorithm, such as Newton-Krylov method \citep{viswanath2007recurrent, willis2013revealing} and adjoint method \citep{farazmand2016adjoint}. Furthermore, any recurrence measure, monitored to find a suitable initial condition for the searching algorithm, {becomes increasingly large} with increased Reynolds number, as turbulence contains increasingly many different scales. {Therefore, a few new approaches to bypass this difficulty have been proposed \citep{page2020,page2022recurrent,parker2022variational, azimi2022constructing}}. Finally, the leading Lyapunov exponent of turbulence grows faster than the inverse of the Kolmogorov time scale \citep{mohan2017scaling}, indicating that finding good initial guesses for a periodic orbit becomes increasingly difficult at higher Reynolds numbers at least with some of the most popular techniques currently available \cite[][]{viswanath2007recurrent, willis2013revealing}.

In earlier studies, these difficulties have been bypassed by modelling some physical processes of turbulence instead of using full DNS. A popular approach taken was to model the small-scale eddies around the large-scale coherent structures of interest using an eddy viscosity implemented by an over-damped large-eddy simulation \citep{rawat2015self, hwang2016invariant, yang2019exact, van2019time,azimi2020self, sasaki2021bifurcation}. However, this approach eliminates the complex nonlinear interactions between the large-scale coherent structures and background turbulence, often offering only limited information on the physical process involved (e.g. self-sustaining process). To this end, in the present study, a step forward from the early approach will be taken. In particular, the objective of the present study is to explore the state space of a reduced-order model (ROM) for the minimal multi-scale (i.e. two-scale) flow unit in plane Couette flow with the computation of its invariant solutions. We will see that the ROM-based approach enables us to delicately control the degrees of freedom of the flow to be sufficiently low (only 600 degrees of freedom), while retaining all the multi-scale physical processes of interest. This significantly relieves the difficulties in the computation of invariant solutions, allowing us to compute a large number of equilibria and unstable periodic orbits. 

The ROM used in the present study was recently proposed by \citet{Cavalieri-ROM}, who demonstrated that it exhibits long-time numerical stability with reasonable statistical agreement to reference DNS.  A classical approach to formulate a ROM is based on the Galerkin projection of the POD (proper orthogonal decomposition) modes, first introduced in the context of turbulence by \citet{lumley1967structure}.  In this approach, the POD basis functions, representing coherent structures, are ranked by their turbulent kinetic energy, allowing the highest energy-containing structures to be included in the model. The approach has been used extensively in a wide variety of applications in fluid dynamics using both experimental and numerical data \cite[e.g.][and many others]{aubry1988dynamics,noack2003hierarchy,smith2005low,khoo2022sparse}. Such an approach, however, requires a priori DNS data to construct the model, and thus it can be computationally demanding. A benefit of the approach by \citet{Cavalieri-ROM} is that the orthogonal basis functions considered are the POD modes of the stochastic response of the linearised Navier-Stokes equations, as such basis functions can be obtained a priori by solving the related Lyapunov equation \citep{farrell1993stochastic, bamieh1999disturbance, jovanovic_bamieh_2005} at minimal computational cost. 

{The ROM of \citet{Cavalieri-ROM} is designed for plane Couette flow, but the multi-scale nonlinear processes of interest in this study were discovered from the shear stress-driven flow model of \cite{doohan_willis_hwang_2021}. Given the only difference between the two systems is the top boundary condition, we expect that the multi-scale processes of interest would be well captured by the ROM of \citet{Cavalieri-ROM} -- indeed, we shall see that this ROM retains all the multi-scale physical processes discovered by \cite{doohan_willis_hwang_2021}, thereby making the dynamics of its invariant solutions physically meaningful. Furthermore, plane Couette flow is a flow configuration where a large number of equilibria and periodic orbits previously computed are well documented \cite[e.g.][]{gibson2009equilibrium,cvitanovic2010geometry}. As will be evident (see also \S\ref{sec:conclusion}), the ROM of \citet{Cavalieri-ROM} also has the potential to be used for the improvement of existing search methods for equilibria and periodic orbits. From this perspective, retaining the plane Couette flow setting in the ROM would also be beneficial.}

This paper is organised as follows: \S \ref{sec:ROM-form} summarises the formulation of the ROM, \S \ref{sec:Energy_Budget_Chp} discusses its validation with respect to a time-averaged energy budget as well as the temporal dynamics of spatially-averaged relevant observables, \S \ref{sec:EQ-PO} discusses the computation of the invariant solutions and discusses the corresponding results, followed by a conclusion in \S \ref{sec:conclusion}. Additional details about the terms in the energy budget equations are contained in Appendix \ref{appA}, and further results discussing the validity of the ROM are contained in the supplementary material.

\section{Reduced-order model} \label{sec:ROM-form}
\subsection{Model formulation}
Incompressible plane Couette flow is considered, where two infinitely wide and long parallel plates, separated by a wall-normal distance $2h$, are set to move in opposite directions with velocity $\pm U_0$. All variables are made dimensionless with length and velocity scales $h$ and $U_0$ respectively. The dimensionless flow domain of interest is considered to be $\boldsymbol{x}(=(x,y,z)) \in [0,2\upi] \times [-1,1] \times [0,\upi]$, equipped with periodic boundary conditions in the streamwise $(x)$, and spanwise $(z)$ directions and the velocity $\boldsymbol{\Tilde{u}}\,|_{y=\pm1} = (\pm1,0,0)$ specified on the upper and lower wall-normal $(y)$ boundaries.  The velocity field is first decomposed into the laminar state, $\boldsymbol{u}_0=(y,0,0)$, and the fluctuations about this state, $\boldsymbol{u}'$:
 \begin{equation}
     \boldsymbol{\Tilde{u}}(\boldsymbol{x},t) = \boldsymbol{u}_0(y) + \boldsymbol{u}'(\boldsymbol{x},t).
 \end{equation}
 The evolution of this fluctuating velocity field is described by the following equations,
\begin{equation}
    \partial_t \boldsymbol{u}' + (\boldsymbol{u}_0\bcdot \bnabla) \boldsymbol{u}' + (\boldsymbol{u}'\bcdot \bnabla) \boldsymbol{u}_0 + (\boldsymbol{u}'\bcdot \bnabla) \boldsymbol{u}'= -\bnabla p + \frac{1}{\Rey}\nabla^2\boldsymbol{u}',
    \label{eq:mom-fluc}
\end{equation}
\begin{equation}
    \bnabla \bcdot \boldsymbol{u}' = 0,
\end{equation}
with boundary conditions given by $\boldsymbol{u}'\,|_{y=\pm1} = (0,0,0)$.

Following the formulation of \citet{Cavalieri-ROM}, we define an inner product,
\begin{equation}    
    \langle \boldsymbol{f},\boldsymbol{g} \rangle = \frac{1}{4\upi^2} \int_0^{\upi} \int_{-1}^1 \int_0^{2\upi} \boldsymbol{f}(\boldsymbol{x}) \bcdot \boldsymbol{g}(\boldsymbol{x}) \:\mathrm{d}x \, \mathrm{d}y \, \mathrm{d}z,
    \label{eq:inner-prod}
\end{equation}
and introduce a modal decomposition of the fluctuating velocity field,
\begin{equation}
    \boldsymbol{u}'(\boldsymbol{x},t) = \sum_{j=1}^N a_j(t) \: \boldsymbol{\Phi}_j(\boldsymbol{x}),
    \label{eq:ansatz}
\end{equation}
where the modes $\boldsymbol{\Phi}_j$ are defined to be divergence free $\bnabla \bcdot\boldsymbol{\Phi}_j =0$, and orthonormal with respect to the inner product defined, $\langle \boldsymbol{\Phi}_i,\boldsymbol{\Phi}_j \rangle = \delta_{ij}\,$, where $\delta_{ij}$ is the Kronecker delta. Hence, by construction, any linear superposition of modes satisfies both the incompressibility condition and the boundary conditions of the system. 

Substituting the decomposition (\ref{eq:ansatz}) into (\ref{eq:mom-fluc}) and taking an inner product with the mode $\boldsymbol{\Phi}_i$ leads to a system of ordinary differential equations describing the evolution of the mode coefficients $a_i(t)$: 
\begin{equation}
        \frac{da_i}{dt} = \frac{1}{\Rey} \sum_{j=1}^N L_{ij} \,a_j + \sum_{j=1}^N \Tilde{L}_{ij} \,a_j + \sum_{j=1}^N \sum_{k=1}^N Q_{ijk} \,a_j a_k,
    \label{eq:ROM}
\end{equation}
where
\begin{subeqnarray}
    L_{ij} & = & \langle \nabla^2 \boldsymbol{\Phi}_j,\boldsymbol{\Phi}_i\rangle,\\
    \Tilde{L}_{ij} & = & -\langle (\boldsymbol{\Phi}_j \bcdot \nabla \boldsymbol{u}_0 + \boldsymbol{u}_0\bcdot\nabla \boldsymbol{\Phi}_j)\: , \boldsymbol{\Phi}_i \rangle,\\
    Q_{ijk} & = & -\langle (\boldsymbol{\Phi}_j \bcdot \nabla \boldsymbol{\Phi}_k)\:, \boldsymbol{\Phi}_i \rangle.
\end{subeqnarray}
Here, $L_{ij}$ and $\Tilde{L}_{ij}$ represent the viscous term and the linear interaction with the laminar state respectively, whereas $Q_{ijk}$ represents the quadratic nonlinear interaction between the different modes.  The pressure dependence is seen to vanish due to the divergence-free property of the modes.

This system of ordinary differential equations still describes the exact dynamics in the limit as $N\rightarrow \infty$ by the orthogonality of the modes, although here we attempt to choose the modes in an optimal way as to model the dynamics of the system using a greatly reduced number of degrees of freedom.

Many choices exist for the modes $\boldsymbol{\Phi}_j(\boldsymbol{x})$. In this study, POD modes of the stochastic response of the linearised Navier-Stokes equations are chosen, and the detailed computational procedure is described in \citet{Cavalieri-ROM}, who followed the method of \citet{jovanovic_bamieh_2005}. The velocity covariance resulting from the stochastic response is obtained by formulating the controllability Gramian of the Navier-Stokes equations linearised about the laminar basic state, $\boldsymbol{u}_0$, and is computed by solving {the associated} Lyapunov equation. The POD modes are obtained by computing the eigenfunctions of the velocity covariance that are ranked in terms of their energy (i.e. eigenvalue). 

Compared to a typical POD approach, the use of POD modes from the stochastic response of the linearised Navier-Stokes equations removes the necessity to form these modes using large numerical or experimental datasets, as they can be directly computed from the forced linear system.  The results of \citet{Cavalieri-ROM} suggest that these modes provide an efficient set of basis functions for the construction of the ROM, as the resulting ROM provides equal or more accurate results than a Galerkin-POD approach constructed from DNS data at least in terms of the second-order velocity statistics.

\subsection{Numerical Implementation}
To implement this model numerically, the POD modes from the controllability Gramian are represented using a Fourier discretisation in the streamwise and spanwise directions, and a Chebychev discretisation in the wall-normal direction,
\begin{equation}
    \boldsymbol{\Phi}_j(\boldsymbol{x}) = \boldsymbol{\hat{\Phi}}_j(y)\,e^{\mathrm{i}(m k_{x,0}\,x + n k_{z,0}\,z)},
    \label{eq:discretisation}
\end{equation}
{where $\hat{\cdot}$ denotes the Fourier transform, and $k_{x,0} = {2\pi}/{L_x} = 1$ and $k_{z,0} = {2\pi}/{L_z} = 2$ denote the fundamental wavenumbers in the streamwise and spanwise directions respectively.}  In particular, we use the modes constructed in \citet{Cavalieri-ROM} with the number of modes $N=600$: $m=\{0,1,2\}, n=\{-2,-1,0,1,2\}$ with $N_{y}=24$, where $N_{y}$ is the number of POD modes used in the expansion for a given $m$ and $n$. Here, we note that only positive streamwise wavenumbers need be considered since negative wavenumber modes can be reconstructed by complex conjugation. {The number of degrees of freedom here was chosen by running a preliminary test, and $N=600$ was found to be close to the smallest possible dimension for a system retaining self-sustaining processes at two scales. Two wavenumbers must be considered for each of the two wall parallel directions to retain the `two' scales, and $N_y=24$ was chosen to ensure the linear stability of the laminar solution. A bifurcation analysis of the ROM was subsequently carried out revealing that its behaviour is remarkably similar to that of plane Couette flow: i.e. saddle-node bifurcation and the resulting bifurcation cascade leading to turbulence \cite[e.g.][]{Kreilos_Eckhardt2012}.} These modes can then be used to construct the system of ordinary differential equations (\ref{eq:ROM}) which are integrated in time using a standard 4th- or 5th-order Runge-Kutta scheme at $\Rey = 1200$ using the \texttt{ode45} function in MATLAB. 

\begin{figure}
  \centerline{\includegraphics[width = 11cm]{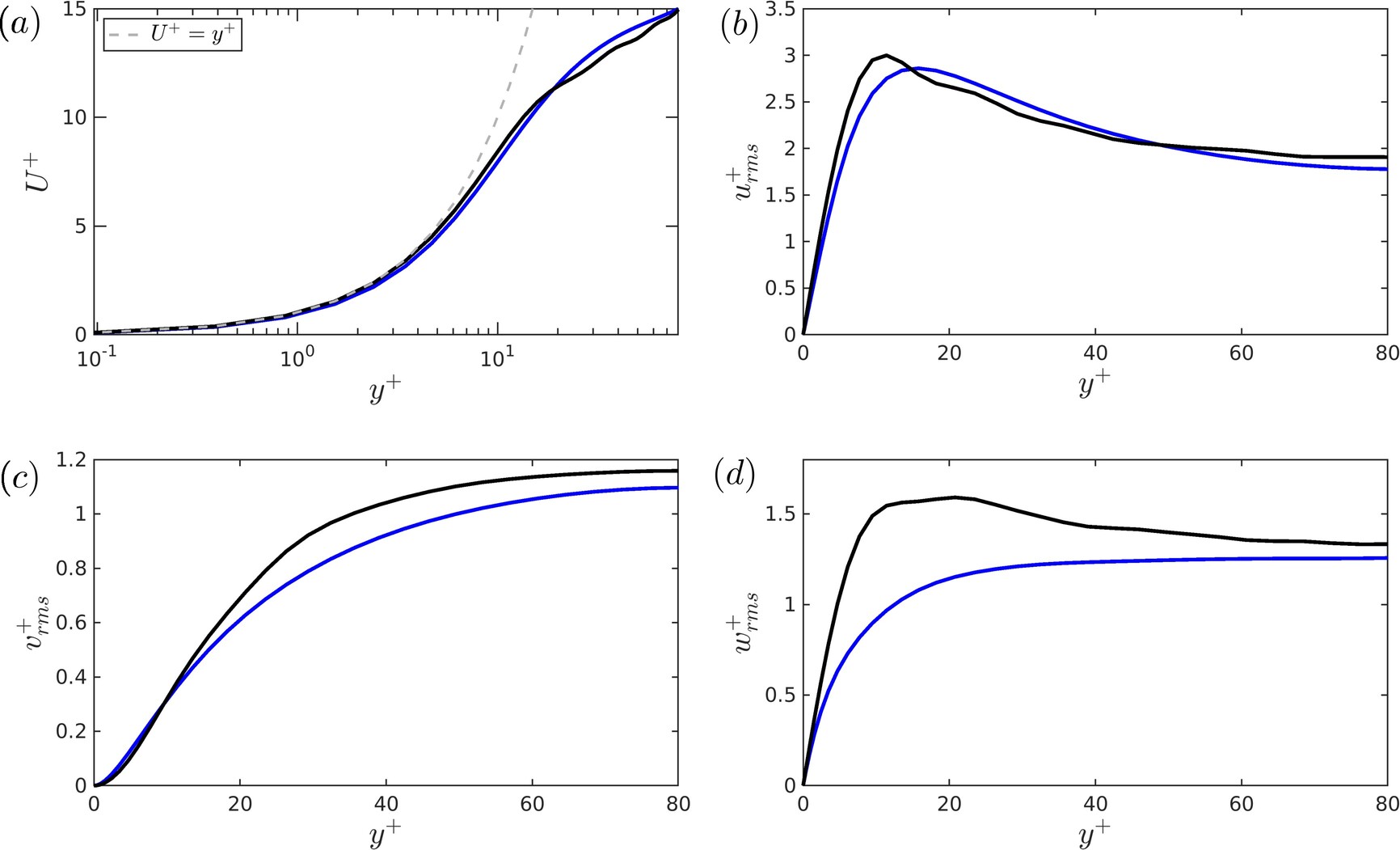}}
  \caption{Mean and rms velocity profiles: $(a)$ $U^+$; $(b)$ $u_{rms}^+$; $(c)$ $v_{rms}^+$; $(d)$ $w_{rms}^+$. The superscript $(\cdot)^+$ denotes the viscous inner-scaled variables. Here, the black and blue lines indicate those from the ROM and DNS, respectively.}
\label{fig:mean-profile}
\end{figure}

The time evolution of this system starting from a random initial condition displays chaotic behaviour consistent with turbulence, as shown by \citet{Cavalieri-ROM}. 
{The long-trajectory was initialised from a random perturbation of the mode coefficients $a_i$ with the amplitude between 0 and 0.1, and was time integrated using \texttt{ode45} with a relative error tolerance of $10^{-8}$.}
Temporal averaging of the flow snapshots from the model over a sufficiently long time leads to the mean and root mean square (rms) velocity profiles shown in figure \ref{fig:mean-profile}. 
These profiles have been compared to reference DNS statistics at a grid resolution of $(N_x, N_y, N_z)=(64, 65, 64)$ computed using an open-source code Channelflow \cite[][{\small\texttt{http://channelflow.org/}}]{channelflow}, showing reasonable agreement relative to the greatly reduced computational cost. {Note that the number of degrees of freedom of DNS is $N_{DNS} \approx 10^6$.} Computation of the velocity gradient at the wall leads to a friction Reynolds number $\Rey_\tau \approx 80$, giving a domain size of $(L_x^+, L_y^+, L_z^+) = (504,160,250)$ based on inner units. These dimensions are bigger than the minimal flow unit configurations of \cite{hamilton_kim_waleffe_1995} and \citet{kawahara_kida_2001}, and are of similar size to that of the minimal multi-scale flow unit formulated by \citet{doohan_willis_hwang_2019}.
In this regard, we expect similar two-scale behaviour as observed by \citet{doohan_willis_hwang_2019} to be captured by the ROM.

\section{Energy budget and dynamics} \label{sec:Energy_Budget_Chp}
To ensure that any invariant solutions found for the ROM are physically relevant to the corresponding exact flow, we first examine the two-scale time-averaged energy budget of the flow as a function of the wall-normal direction.  The results of this energy budget can then be qualitatively compared to the results of the energy budget analysis of the similar two-scale system studied by \citet{doohan_willis_hwang_2021}.

\subsection{Two-scale energy budget}
To formulate a relevant energy budget for a statistically stationary turbulent state, the full velocity field is decomposed as
\begin{equation}
    \boldsymbol{\Tilde{u}}(\boldsymbol{x},t) = \boldsymbol{U}(y) + \boldsymbol{u}(\boldsymbol{x},t),
\end{equation}
where $\boldsymbol{U}$ describes the mean flow, defined as $\boldsymbol{U}(\boldsymbol{x}) = \frac{1}{T}\int_0^T \boldsymbol{\Tilde{u}}(\boldsymbol{x},t)\,\mathrm{d}t$ for some suitably long time interval $T$, and $\boldsymbol{u}(\boldsymbol{x},t)$ is the velocity fluctuations about the mean flow.

As discussed in \cite{doohan_willis_hwang_2019}, the inner-scaled flow domain at $Re_\tau(\simeq 80)$ considered here is about $L_z^+\simeq 250$. Given that the smallest spanwise size of energy-containing structures is $\lambda_z^+\simeq 100$, the computational domain would admit energy-containing structures mainly at two different spanwise length scales: $\lambda_z^+\simeq 125, 250$. To understand the interactions of the two different scales within the flow, we further decompose the fluctuating velocity field into a large- and small-scale fluctuating velocity field,
\begin{equation}
    \boldsymbol{{u}}(\boldsymbol{x},t) = \boldsymbol{{u}}_l(\boldsymbol{x},t) + \boldsymbol{u}_s(\boldsymbol{x},t),
\end{equation}
where the large- and small-scale velocity fields are defined through the large- and small-scale projection operators, $\mathcal{P}_l\{\cdot\}$ and $\mathcal{P}_s\{\cdot\}$, which split the fields by spanwise wavenumber in the following way:
\begin{subequations}
\label{eq:velocity-ls}
\begin{alignat}{3}
 \boldsymbol{u}_l(\boldsymbol{x},t) &= \mathcal{P}_l\{\boldsymbol{u}(\boldsymbol{x},t)\} &&= \sum_{|m|\le m_x}\: \: \sum_{|n|\le 1} &&\boldsymbol{\hat{u}}(y)\: e^{\mathrm{i}(m k_{x,0}\,x + n k_{z,0}\,z)},\\ 
\boldsymbol{u}_s(\boldsymbol{x},t) &= \mathcal{P}_s\{\boldsymbol{u}(\boldsymbol{x},t)\} &&=\sum_{|m|\le m_x} \sum_{2\le|n|\le n_z}&& \boldsymbol{\hat{u}}(y)\: e^{\mathrm{i}(m k_{x,0}\,x + n k_{z,0}\,z)}. 
\end{alignat}
\end{subequations}

The split by spanwise wavenumber here is justified by the observation that the statistical structure of eddies in wall-bounded flows is self-similar with respect to the spanwise wavelength and proportional to the distance from the wall \citep{hwang_2015}.  Additionally, both large and small scales are expected to contain a range of streamwise wavenumbers due to the presence of both long streaks and short streamwise vortices as part of the self-sustaining process \citep{hwang_2015}.

The evolution equations for turbulent fluctuations are subsequently projected onto the large- and small-scale velocity subspaces using the definition of $\mathcal{P}_l\{\cdot\}$ and $\mathcal{P}_s\{\cdot\}$, yielding two fluctuation equations, one equation each for the large and small scales respectively.  Multiplying each of these equations by the corresponding large- or small-scale velocity and averaging in the streamwise and spanwise direction yields energy budget equations for each component of the velocity at large,
\begin{subequations}
\label{eq:energy-eq-large}
\begin{align}
& \frac{\partial E_{ul}}{\partial t} = {P}_{ul} + T_{ul} + \Pi_{ul} + T_{\nu,ul} + \varepsilon_{ul}, \\
& \frac{\partial E_{vl}}{\partial t} = T_{vl} + \Pi_{vl} + T_{\nu,vl} + T_{p,vl} + \varepsilon_{vl}, \\
& \frac{\partial E_{wl}}{\partial t} = T_{wl} + \Pi_{wl} + T_{\nu,wl} + \varepsilon_{wl},
\end{align}
\end{subequations}
and small scales,
\begin{subequations}
\label{eq:energy-eq-small}
\begin{align}
& \frac{\partial E_{us}}{\partial t} = {P}_{us} + T_{us} + \Pi_{us} + T_{\nu,us} + \varepsilon_{us}, \\
& \frac{\partial E_{vs}}{\partial t} = T_{vs} + \Pi_{vs} + T_{\nu,vs} + T_{p,vs} + \varepsilon_{vs}, \\
& \frac{\partial E_{ws}}{\partial t} = T_{ws} + \Pi_{ws} + T_{\nu,ws} + \varepsilon_{ws},
\end{align}
\end{subequations}
where $P$ denotes the energy production, $T$ the turbulent transport, $\Pi$ the pressure strain, $T_\nu$ the viscous transport, $T_p$ the pressure transport, and $\epsilon$ the dissipation at each scale and component, where the subscript $l$ and $s$ denote large and small scales, and $u$, $v$ and $w$ streamwise, wall-normal and spanwise components. Full details of each term are included in Appendix \ref{appA}.  Since only the velocity field is computed directly by the ROM, the fluctuating pressure field must be computed using the Poisson equation for pressure projected onto the large and small scales:
\begin{subequations}
\begin{align}
    & \nabla^2 p_l = - U\frac{\partial^2 u_l}{\partial x^2} - \frac{\partial v_l}{\partial x} \frac{\partial U}{\partial y}  - \mathcal{P}_l\{\bnabla \bcdot (\boldsymbol{u}\bcdot \bnabla\boldsymbol{u})\}, \\
    & \nabla^2 p_s = - U\frac{\partial^2 u_s}{\partial x^2} - \frac{\partial v_s}{\partial x} \frac{\partial U}{\partial y}  - \mathcal{P}_s\{\bnabla \bcdot (\boldsymbol{u}\bcdot \bnabla\boldsymbol{u})\},
\end{align}
\label{eq:Poisson}
\end{subequations}
with boundary conditions specified as $\partial_y\, p_l |_{y=\pm1}= 0$, and $\partial_y\, p_s |_{y=\pm1}= 0$ \citep{kim_1989}.

\begin{figure}
  \centerline{\includegraphics[width = 13cm]{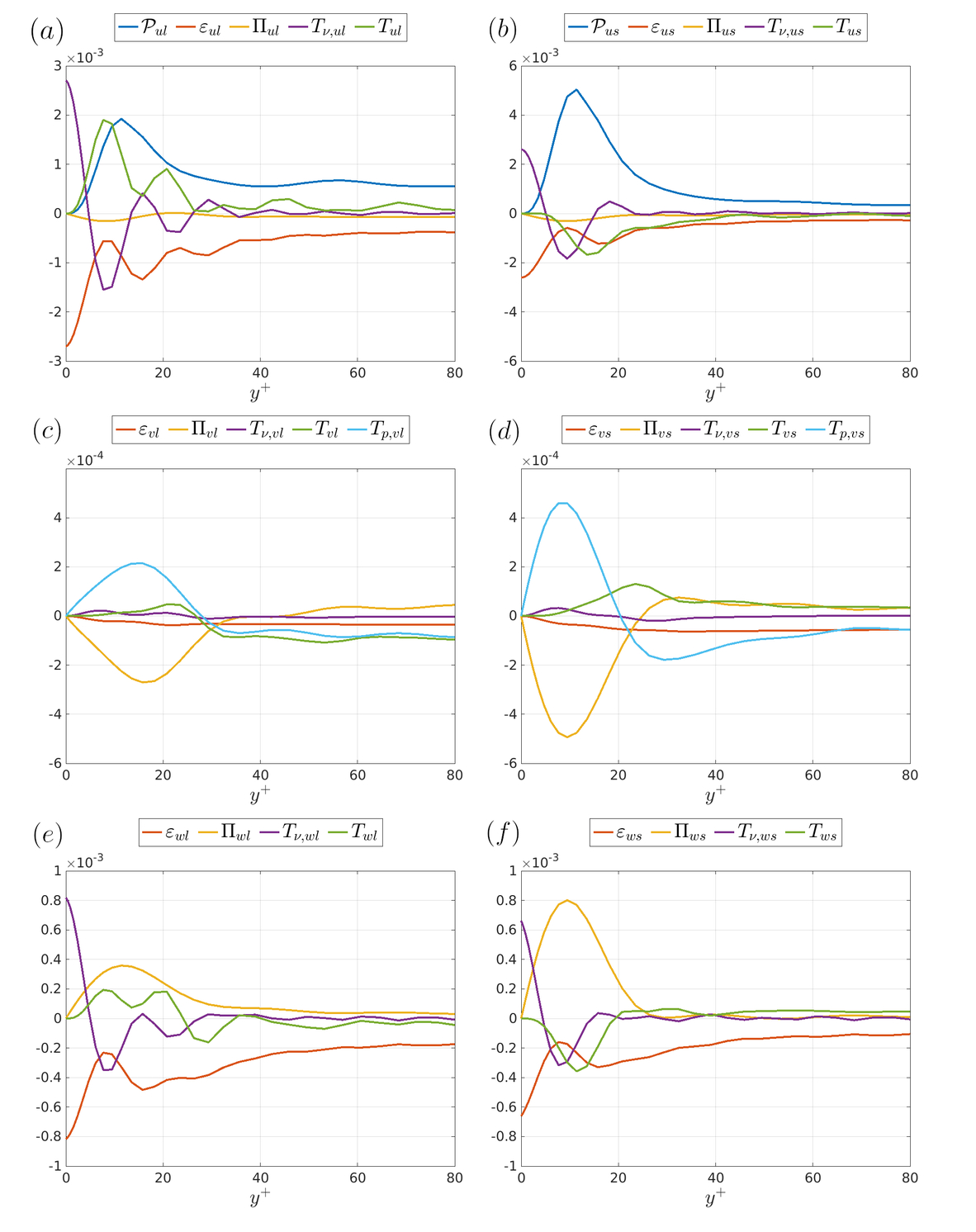}}
  \caption{Wall-normal variation of the time-averaged terms of the energy budget equations (\ref{eq:energy-eq-large}) and (\ref{eq:energy-eq-small}) for $(a,b)$ streamwise, $(c,d)$ wall-normal and $(e,f)$ spanwise component: $(a,c,e)$ large scale; $(b,d,f)$ small scale. }
\label{fig:budget}
\end{figure}

The various multi-scale energy budget terms have been computed from the reduced-order model. The temporal averaging has been taken over a dataset which is roughly 18,000 dimensionless time units in length $(T^+ = T u_\tau^2 /\nu \approx 96,500)$. They have been plotted over a channel half-height as a function of the wall-normal direction in figure \ref{fig:budget}. The overall statistical behaviour of the two-scale energy budget terms are qualitatively very similar to those found by \citet{doohan_willis_hwang_2021} who used the DNS results of their two-scale shear stress-driven flow model. 

A large amount of energy production is seen in both the large and small scales (blue lines in figures \ref{fig:budget}$a,b$), featuring peaks in production in the near-wall region at approximately $y^+\approx 10$, the location of which is consistent with that reported by \citet{Production-peak1} and \citet{Production-peak2} as well as many others. The magnitude of peak of the small scale production is seen to be larger than that of the large scales, similar to the results reported by \citet{doohan_willis_hwang_2021}, although slightly less production is observed overall in the logarithmic region of the flow compared to this work.

The pressure strain terms (yellow lines in figure \ref{fig:budget}) sum to zero at each scale as a consequence of continuity, i.e. $\Pi_{ul}+\Pi_{vl}+\Pi_{wl}=0$ and $\Pi_{us}+\Pi_{vs}+\Pi_{ws}=0$.  The streamwise components of the pressure strain are negative across the whole wall-normal domain, whereas the wall-normal and spanwise pressure strains are seen to be positive, except $y^+ \lesssim 20$. This is indicative of the pressure strain terms being responsible for the redistribution of the turbulent kinetic energy from the streamwise to the wall-normal and spanwise components, also observed in \citet{doohan_willis_hwang_2021}.  However, the pressure strain terms have a relative magnitude lower than that of \citet{doohan_willis_hwang_2021}, presumably due to the low number of streamwise and spanwise Fourier modes used for the present ROM. For $y^+ \lesssim 20$, large peaks in the spanwise and wall-normal components are observed. The near-wall negative wall-normal pressure strain is attributed to the well-known `splat effect' \cite[]{mansour_kim_moin_1988,perot_moin_1995,lee_moser_2019}, whereby fluid motions moving towards the wall must be turned to move parallel to the wall due to its impermeability.

The streamwise turbulent transport term at the large scales is seen to be positive, especially in the near-wall region (green line in figure \ref{fig:budget}$a$), implying that the turbulent transport from the small scales injects energy into the large scales. This is the inverse energy transfer (or energy feeding from the small scales), previously observed \cite[]{cho_hwang_choi_2018,doohan_willis_hwang_2021}, and it gradually disappears on increasing $y^+$. In general, the turbulent transport terms follow similar trends to that of \citet{doohan_willis_hwang_2021} but are more oscillatory  (green lines in figure \ref{fig:budget}). This lack of smoothness is likely a result of the nonlinear interactions between the truncated modes not being fully captured by the model, although the consequences of this are not immediately clear.  For this reason, the turbulent transport terms have been further decomposed and studied in \S\ref{sec:energy-transport} as their role is of significant interest in relation to the accuracy of the multi-scale behaviour in ROM.

The largest amounts of dissipation are seen in the near-wall region through the streamwise component (orange line in figure \ref{fig:budget}$a$), similar to \citet{doohan_willis_hwang_2021}. There is slightly greater large-scale dissipation than small-scale dissipation for all components (orange lines in figure \ref{fig:budget}), suggesting that the cascade of energy from large to small scales \citep{kolmogorov1941local} is not fully resolved. This is likely due to the truncated nature of the model, whereby energy cannot be transferred to the smaller scales neglected by the model, as a result of spectral blocking. Nevertheless, the amounts of dissipation at each scale are still comparable due to the low $Re$ considered here, as in \citet{doohan_willis_hwang_2021}. {However, at higher Reynolds numbers, the energy cascade is expected to be more poorly resolved due to the increasing number of small-scale eddies for turbulence dissipation. As such, the number of degrees of freedom of the ROM would need to be increased further. }

Lastly, the pressure and viscous transport terms appropriately balance the other terms, and their behaviours are also consistent with those found in \cite{doohan_willis_hwang_2019}. The integrated energy is seen to be balanced over all terms to within 2\% relative to the magnitude of the turbulent production.

\subsection{Inter- and intra-scale energy transport} \label{sec:energy-transport}
To access the role of the turbulent transport in greater detail, we decompose these terms further, following the methods of \citet{doohan_willis_hwang_2021} and \citet{kawata_alfredsson_2019}:
\begin{subequations} \label{eq:transport-decomposition}
\begin{alignat}{3}
& T_{ul} = T_{ul,-} &&+T_{ul,\#} &&\: -T_{u,\updownarrow},\\
& T_{vl} = T_{vl,-} &&+T_{vl,\#} &&\: -T_{v,\updownarrow},\\
& T_{wl} = T_{wl,-} &&+T_{wl,\#} &&\: -T_{w,\updownarrow},\\
& T_{us} = T_{us,-} &&+T_{us,\#} &&\: +T_{u,\updownarrow},\\
& T_{vs} = T_{vs,-} &&+T_{vs,\#} &&\: +T_{v,\updownarrow},\\
& T_{ws} = T_{ws,-} &&+T_{ws,\#} &&\: +T_{w,\updownarrow}.
\end{alignat}
\end{subequations}
Here, $T_-$ denotes the intra-scale spatial turbulent transport, $T_\#$ the inter-scale spatial turbulent transport, and $T_\updownarrow$ the inter-scale turbulent transport. Here the use of \emph{intra} refers to the involvement of velocity fields of only the same scale, whereas \emph{inter} refers to terms that include the nonlinear coupling of velocity fields of different scales.  \emph{Spatial} transport refers to terms that can be rewritten as the divergence of a vector field, meaning that they do not contribute to the loss or gain of energy when integrated over the spatial domain. On the contrary, the \emph{inter-scale turbulent} transport cannot be written as the divergence of a vector field and is thus directly responsible for energy exchanges between the large and small scales. Importantly, these inter-scale turbulent transport terms appear in both the large- and small-scale energy transport equations with opposing signs, thus cancelling each other out when considering the total energy budget of both the large and small scales.  The full details of these equations are included in Appendix \ref{appA}. 

\begin{figure}
  \centerline{\includegraphics[width = 13cm]{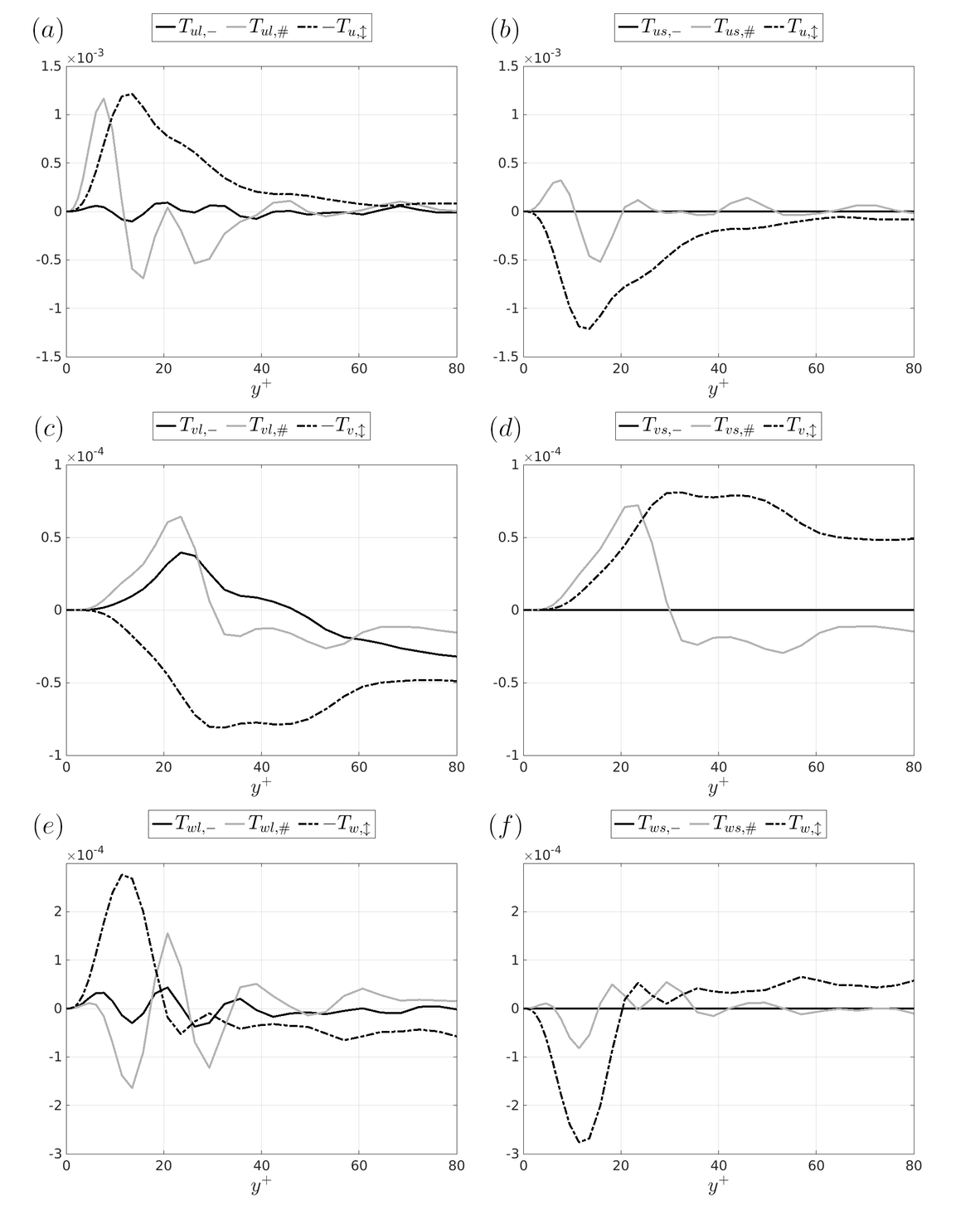}}
  \caption{Wall-normal variation of the time-averaged intra- and inter-scale energy transport terms in (\ref{eq:transport-decomposition}) for $(a,b)$ streamwise, $(c,d)$ wall-normal and $(e,f)$ spanwise components: $(a,c,e)$ large scale; $(b,d,f)$ small scale.}
\label{fig:transport}
\end{figure}

The time-averaged energy transport terms in (\ref{eq:transport-decomposition}) are plotted as a function of the wall-normal direction in figure \ref{fig:transport}. The inter-scale turbulent transport terms, $T_{\updownarrow}$, are responsible for the direct transport of energy between scales.  In the streamwise component, only inter-scale turbulent transport, $T_{\updownarrow}$, from the small to large scales is observed across the entire domain in the mean (figures \ref{fig:transport}$a,b$).  This is contrary to the DNS results of \citet{doohan_willis_hwang_2021}, which show a contribution to the energy cascade in the outer region of the flow in the streamwise direction. With that being said, the spanwise and wall-normal components are seen to contribute to the energy cascade from large to small scale in a similar way to that of \citet{doohan_willis_hwang_2021}, suggesting that the cascade is still being captured, at least to some degree, in the outer region of the flow (figure \ref{fig:transport}$c$-$f$). The streamwise and spanwise components of the inter-scale turbulent transport, $T_{\updownarrow}$, show a large amount of energy transport from the small to large scales in the near wall region (figure \ref{fig:transport}).  This is representative of the near-wall energy feeding process from small to large scale, consistent with both \citet{cho_hwang_choi_2018} and \citet{doohan_willis_hwang_2021}. {Note that this inverse energy transfer from small to large scale \cite[e.g.][]{cho_hwang_choi_2018,Andreolli2021} has been known to break the self-similarity of the energy-containing motions in the region close to the wall, where the attached eddy hypothesis is not supposed to be valid \cite[for a further discussion, see][]{Hwang2016c}}.

The inter-scale spatial transport, $T_{\#}$, describes an energy transport that results from interactions between the different scales. The streamwise inter-scale spatial transport terms show local maxima at a wall-normal location similar to that of the peaks of the inter-scale turbulent transport and production, suggesting that these transport terms also play a role in transporting the turbulent kinetic energy produced by the flow between the scales.  However, all of these inter-scale spatial transport terms exhibit similar trends at both the large and small scales (at least with respect to their signs) to those in \citet{doohan_willis_hwang_2021}, suggesting that these terms act in a similar manner at both scales.

The intra-scale spatial transport, $T_{-}$, represents energy transport within each scale.  In the large scales, these terms follow reasonably similar trends to \citet{doohan_willis_hwang_2021}. 
However, no intra-scale transport is exhibited by the ROM at the small scales, a feature that is not in line with the DNS results of \citet{doohan_willis_hwang_2021}.  This is a consequence of spanwise averaging over the small-scale intra-scale transport terms equalling zero. For example, the streamwise intra-scale spatial turbulent transport is given by 
\begin{equation} \label{eq:triple}
     T_{us,-} = -\langle  u_s (\boldsymbol{u}_s\bcdot \nabla u_s) \rangle_{x,z}.
\end{equation}
Given that $T_{us,-}$ takes the form $\langle f(x,y,n,t) e^{\mathrm{i}2(n_1+n_2+n_3)z}\rangle_{x,z}$, where $n_i$ refers to the spanwise wavenumber related to the $i^{th}$ term in the triple correlation of (\ref{eq:triple}).  Since $n_1+n_2+n_3 \neq 0$ in the small scales by account of the model's truncation (i.e. $n_i=\pm2$ from (\ref{eq:velocity-ls})), this spanwise average will always be zero in the small scales unless additional wavenumbers are considered in the present ROM. 

The immediate effect of this on the results of the model is not apparent, but with that being said, intra-scale processes that are responsible for the transport of energy from the production to the dissipation are seen to persist with the pressure strain being largely responsible for this redistribution.  Additionally, the viscous transport, $T_\nu$, which itself only involves intra-scale terms, is seen to be active in relation to the production and dissipation terms at both scales. 

\subsection{Temporal dynamics} \label{sec:energy-dynamics}
Having studied the time-averaged variation of the different terms in the energy equations, it is also of interest to study the temporal relationship between these quantities and to confirm the presence of known dynamic processes found in the similar system of \citet{doohan_willis_hwang_2021}.  These include the large- and small-scale self-sustaining process, the energy cascade, the \emph{driving} mechanism which transports energy from large-scale processes and injects it into the small scales through the small-scale production, and the \emph{feeding} mechanism which is responsible for transporting energy from small to large scales.  To study the temporal correlation of such quantities, we introduce the temporal cross-correlation of two time-varying functions $f(t)$ and $g(t)$,
\begin{equation}
    \{f\star g\}(\tau) = \frac{\int_{\Delta T} f(t+\tau)g(t)\,\mathrm{d}t}{\left( \int_{\Delta T} f(t)^2\,\mathrm{d}t \right) ^{1/2} \left( \int_{\Delta T} g(t)^2\,\mathrm{d}t \right) ^{1/2}},
\end{equation}
where $\Delta T$ is a suitable time interval of interest.

Of particular interest are the presence of the large- and small-scale self-sustaining processes \citep{hwang_bengana_2016,doohan_willis_hwang_2021}.  With this in mind, the energy related to the different velocity components is separated into streamwise dependent and independent parts \citep{lucas_kerswell_2017,doohan_willis_hwang_2019,doohan_willis_hwang_2021}.  Specifically, these provide observables based on the kinetic energy for the structural components of the self-sustaining process: i.e. straight streaks $(E_{ss})$, wavy streaks $(E_{ws})$, straight rolls $(E_{sr})$, and wavy rolls $(E_{wr})$ at both the large,
\begin{subequations}
\begin{alignat}{2}
& E_{ss,l}(t) &&= \tfrac{1}{2} \big\langle \langle u_l \rangle_x^2 \big\rangle _z \big|_0^{80} ,\\
& E_{ws,l}(t) &&= \tfrac{1}{2} \big\langle \big(u_l - \langle u_l \rangle_x\big)^2 \big\rangle _{x,z} \big|_0^{80},  \\
& E_{sr,l}(t) &&= \tfrac{1}{2} \big\langle \langle v_l \rangle_x^2 + \langle w_l \rangle_x^2 \big\rangle _z \big|_0^{80},  \\
& E_{wr,l}(t) &&= \tfrac{1}{2} \big\langle \big(v_l - \langle v_l \rangle_x\big)^2 + \big(w_l - \langle w_l \rangle_x\big)^2 \big\rangle _{x,z} \big|_0^{80}, 
\end{alignat}
\end{subequations}
and small scales,
\begin{subequations}
\begin{alignat}{2}
& E_{ss,s}(t) &&= \tfrac{1}{2} \big\langle \langle u_s \rangle_x^2 \big\rangle _z \big|_0^{40}, \\
& E_{ws,s}(t) &&= \tfrac{1}{2} \big\langle \big(u_s - \langle u_s \rangle_x\big)^2 \big\rangle _{x,z} \big|_0^{40},  \\
& E_{sr,s}(t) &&= \tfrac{1}{2} \big\langle \langle v_s \rangle_x^2 + \langle w_s \rangle_x^2 \big\rangle _z \big|_0^{40},  \\
& E_{wr,s}(t) &&= \tfrac{1}{2} \big\langle \big(v_s - \langle v_s \rangle_x\big)^2 + \big(w_s - \langle w_s \rangle_x\big)^2 \big\rangle _{x,z} \big|_0^{40}, 
\end{alignat}
\end{subequations}
where the shorthand $f(t)|_{y_1}^{y_2}$ denotes averaging of quantity $f(t)$ in the wall-normal direction over the interval $[y_1, y_2]$ given in wall units. The choice of $y^+=40$ to separate the boundary of the near-wall region (viscous sublayer and buffer layer) from the logarithmic region is chosen to approximately separate phenomenological differences observed in these regions based on figures \ref{fig:mean-profile}, \ref{fig:budget} and \ref{fig:transport}.  In practice, small changes in the choice of this separating $y^+$ value was seen to have little impact on the results of the following sections.

\begin{figure}
  \centerline{\includegraphics[width = 13cm]{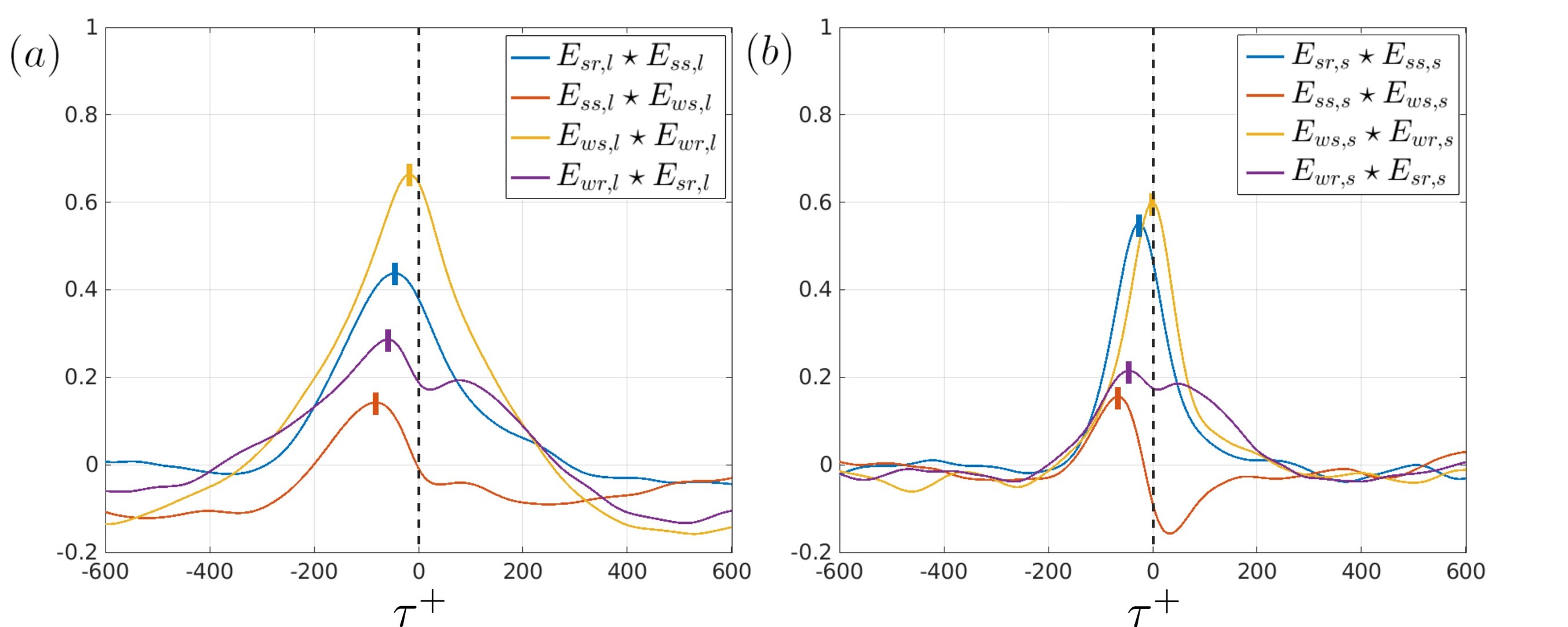}}
  \caption{Temporal cross-correlation between the energy observables associated with the structural components of the self-sustaining process (SSP) at $(a)$ large and ($b$) small scales.  Markers show the peak of each cross-correlation.}
\label{fig:ssp1}
\end{figure}

The cross-correlation of these energies at both scales has been computed and is shown in figure \ref{fig:ssp1}.  Indeed, correlation does exist between the different kinetic energies, with the shift of the peak of the correlations revealing the order of the processes: $E_{sr} \rightarrow E_{ss} \rightarrow E_{ws} \rightarrow E_{wr} \rightarrow E_{sr}\,$, consistent with the results of \citet{hamilton_kim_waleffe_1995}, \citet{hwang_bengana_2016} and \citet{doohan_willis_hwang_2021}, indicating that the self-sustaining process at both scales is captured by the ROM.  Further, the timescale of these processes both in the large and small scales is also similar to that found by \citet{doohan_willis_hwang_2021}.  The small-scale process occurs on a timescale that is significantly shorter than that of its large-scale counterpart, as seen by the width of the cross-correlation functions.

Similar correlations have been produced for the other energy transport processes identified by \citet{doohan_willis_hwang_2021}, namely for the energy cascade, driving, and feeding processes, all showing agreement with those found in the original DNS.  The details of these correlations can be found in the supplemental material of this paper.  In summary, this system contains an energy cascade which transports energy from the large-scale production to small-scale wall-normal and spanwise dissipation mainly through a particular part of the wall-normal inter-scale turbulent transport,
\begin{equation} \label{eq:cascade-decomposition}
    T_{v,\updownarrow c} = \langle  v_l(\boldsymbol{u}_l\bcdot \nabla v_s) \rangle_{x,z}.
\end{equation}
Energy transported from large to small scales also involves the driving mechanism, the activation process of small-scale production via energy cascade. The related energy transport is well characterised by the other part of the wall-normal inter-scale turbulent transport,
\begin{equation} \label{eq:driving-decomposition}
    T_{v,\updownarrow d} = -\langle  v_s(\boldsymbol{u}_s\bcdot \nabla v_l) \rangle_{x,z}.
\end{equation}
It results in subsequent activation of the small-scale production via the Orr mechanism \cite[]{Orr}, promoting a range of other processes such as an increase in small-scale dissipation in response to the breakdown of small-scale wavy streaks as well as a feeding process which transports energy from the small to large scales. Finally, the feeding process is also seen to transport energy from small-scale production mainly to large-scale dissipation terms through the streamwise and spanwise inter-scale turbulent transport, and it involves the activation of small-scale wavy streaks and small-scale wavy rolls (through the spanwise pressure strain).

\section{Equilibria \& periodic orbits} \label{sec:EQ-PO}
Having validated the ROM by ensuring that it captures both the time-averaged energy budget at large and small scales, as well as capturing the known processes associated with the temporal energy dynamics of the flow, we now search for invariant solutions using the model.

Equilibria and periodic orbits were searched using a trust-region algorithm, implemented in the \texttt{fsolve} function in MATLAB. Invariant solutions are steady states of (\ref{eq:ROM}), such that
\begin{equation}
\frac{1}{\Rey} \sum_{j=1}^N L_{ij} \,a_j + \sum_{j=1}^N \Tilde{L}_{ij} \,a_j + \sum_{j=1}^N \sum_{k=1}^N Q_{ijk} \,a_j a_k = 0,
\label{eq:ROMeq}
\end{equation}
whereas periodic orbits are solutions of (\ref{eq:ROM}) satisfying $a_j(t+T) = a_j(t)$, $j=1, \ldots, N$. Equilibria may be sought by looking directly for the zeros of (\ref{eq:ROMeq}). Periodic orbits are searched as zeros of the normalised {residual}
\begin{equation}
r(\mathbf{a}) = \frac{\| \mathbf{a}(t+T) - \mathbf{a}(t) \|}{\| \mathbf{a}(t)\| },
\label{eq:normalised_error}
\end{equation}
where the $\mathbf{a}$ vector collects the temporal coefficients $a_j(t)$ of the ROM.

In order to avoid travelling waves and relative periodic orbits, and thus simplify the search for invariant solutions, the reflection symmetry $[u,v,w](x,y,z) = [-u,-v,-w](-x,-y,-z)$ was imposed in the ROM; this is accomplished by truncating the ROM to the modes that satisfy the said symmetry, resulting in 300 degrees of freedom. This lower dimension simplifies the search for invariant solutions. When looking for equilibria, the analytical Jacobian of (\ref{eq:ROMeq}) was supplied to the Newton solver, which greatly accelerates the search by reducing the number of function calls. Periodic orbits require time integrations to obtain the Jacobian, and using the direct linearisation of the ROM did not lead to clear benefits in our tests; we have thus resorted to a finite-difference approximation of the Jacobian of the normalised error of (\ref{eq:normalised_error}). The tolerance of the solver was specified such that (\ref{eq:ROMeq}) is correct within $10^{-8}$ for equilibria, and that the residual is lower than $10^{-5}$ for periodic orbits -- {the final residuals of many periodic orbits were of $O(10^{-6})-O(10^{-7})$. Lastly, the convergence of time period, which was updated in the Newton iterations, was not explicitly monitored, although an approach for this was recently proposed \cite[]{page2022recurrent}. }

Using this method, starting from initial guesses taken from a recurrence analysis of chaotic trajectories, 96 equilibria, named from EQ1 to EQ96, and 43 periodic orbits, named from PO1 to PO43, have been computed.  The period of the periodic orbits found varies between $T^+ \in [26.53, 141.57]$. In the following sections, the inner units are formulated with respect to the friction velocity of the ROM for each invariant solution and not the friction velocity of each invariant solution unless specified otherwise.  

Equilibria and periodic orbits computed in the similar two-scale system of \citet{doohan_2022-tw} through DNS were not seen to capture the full multi-scale behaviour of the system. They typically captured the dynamics of only a single integral length scale. The solutions that appeared to capture any of the multi-scale dynamics were from the upper branch, i.e. the solution family characterised by a large wall shear stress (i.e. high $\Rey_\tau$). Even still, these upper branch solutions failed to capture the full multi-scale dynamics of the long-time chaotic state.

To compare the solutions of \citet{doohan_2022-tw} to those found in this study using the ROM, we have constructed phase portraits of similar observables, as well as the mean and rms statistics of the invariant solutions, which are presented in the following sections, first for the equilibria and followed by the periodic orbits.  This is followed by a discussion of periodic orbits identified to contain essential dynamics relating to key physical processes in the flow.

\subsection{Equilibria}
\begin{figure}
  \centerline{\includegraphics[width = 11cm]{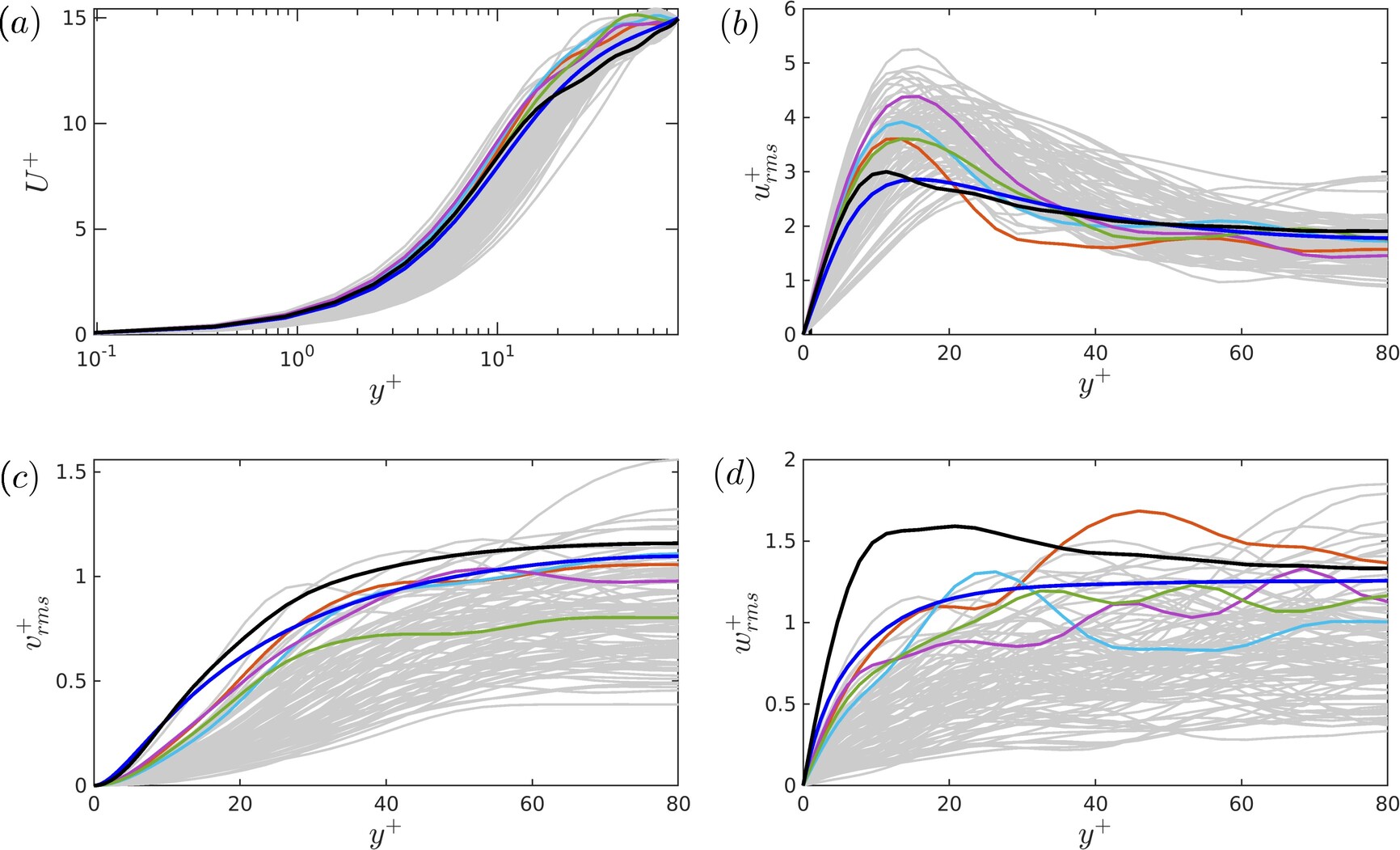}}
  \caption{Mean and root mean square velocity profiles for the equilibrium solutions (grey) compared to the long-time trajectory generated by the ROM (black) and DNS (blue): $(a)$ $U^+$; $(b)$ $u_{rms}^+$; $(c)$ $v_{rms}^+$; $(d)$ $w_{rms}^+$. Highlighted equilibria are EQ8 (orange); EQ28 (cyan); EQ48 (purple); EQ52 (green).}
\label{fig:EQ-statistics}
\end{figure}

\begin{figure}
  \centerline{\includegraphics[width = 13cm]{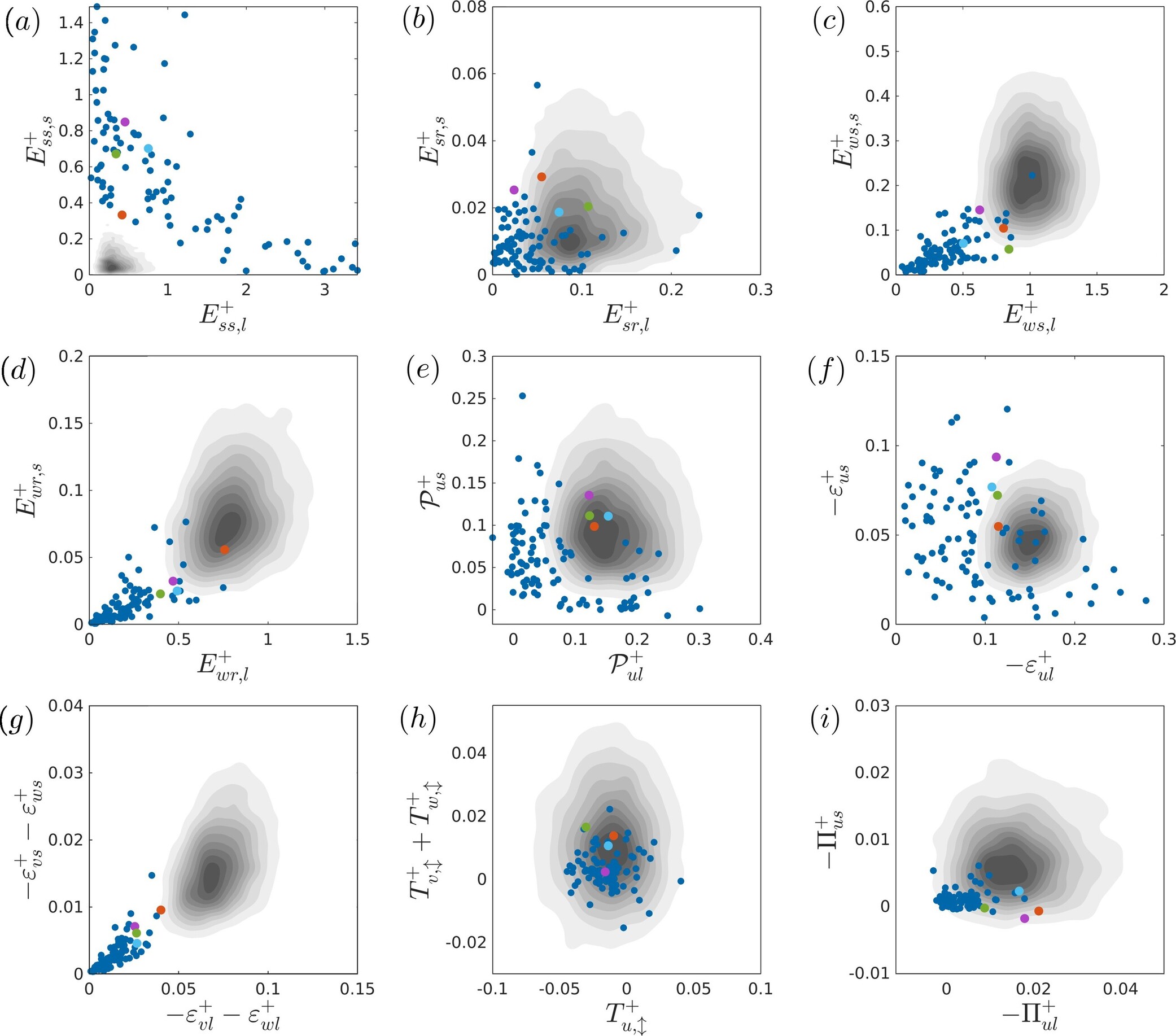}}
  \caption{Phase portraits of the large and small scale energy observables of the self-sustaining process and two-scale energy budget the equilibria (blue), EQ8 (orange), EQ28 (cyan), EQ48 (purple), EQ52 (green), {compared to linearly spaced contours of the respective bivariate probability density function (dark-light grey) } generated from a long time trajectory of the ROM.}
\label{fig:EQ-Energy}
\end{figure}

The mean and rms velocity profiles of the computed equilibrium solutions as functions of the wall-normal direction are shown in figure \ref{fig:EQ-statistics}, compared to a long-time trajectory from the ROM shown in black and the DNS in blue.  Four of these solutions have been highlighted in figure \ref{fig:EQ-statistics} which closely capture the wall shear stress of the long-time trajectory. Phase portraits showing similar energy observables to those of \citet{doohan_2022-tw} are shown in figure \ref{fig:EQ-Energy}, also highlighting the same equilibria as in the velocity profiles. Here, we note that many of the computed equilibria have non-negligible amounts of energy and production at both large and small scales (figures \ref{fig:EQ-Energy}$a$-$d$), and, in this sense, they may be viewed to be multi-scale equilibrium solutions. 

A large number of the equilibria are found to capture the mean flow reasonably well, especially in the near-wall region (figure \ref{fig:EQ-statistics}$a$). However, many did not accurately capture the velocity fluctuations of the flow and are not seen to be highly representative of the chaotic state with respect to observables other than the mean flow. The equilibria found are typically seen to overestimate the near-wall streamwise velocity fluctuations. This is well observed in the maxima of the streamwise rms profiles (figure \ref{fig:EQ-statistics}$b$), which appear to be related to the large amount of energy associated with straight streaks in figure \ref{fig:EQ-Energy}($a$). Equilibria overestimating the amount of energy related to straight streaks compared to the long-time trajectory was also observed by \citet{doohan_2022-tw} with the relative magnitude of the overestimation being similar, especially for the small-scale straight streaks.

Wall-normal fluctuations are seen to be of generally lower magnitude than that of the long-time trajectory, although some of the highlighted equilibria with high friction Reynolds numbers are seen to capture the wall-normal rms velocity profiles quite well (EQ8, EQ28 and EQ48; figure \ref{fig:EQ-statistics}$c$). With an accurate reproduction of the mean wall shear stress, some of the computed equilibria capture the energy production of the large and small scales reasonably well (figure \ref{fig:EQ-Energy}$e$). The spanwise rms velocities of the computed equilibria are also typically lower than that of the chaotic trajectory (figure \ref{fig:EQ-statistics}$d$). These observations are consistent with some of the phase portraits, in which straight rolls, wavy streaks and rolls are not well captured by the equilibria (figures \ref{fig:EQ-Energy}$b$-$d$). 

Finally, the phase portraits associated with energy cascade and nonlinear scale interactions show that most of the equilibria computed do not properly capture the dynamics of the chaotic trajectory. {The streamwise dissipation of many equilibria does not appear near the chaotic trajectory (figure \ref{fig:EQ-Energy}$f$). Furthermore, the phase portraits of the wall-normal and spanwise dissipation (figure \ref{fig:EQ-Energy}$g$), inter-scale energy transport (figure \ref{fig:EQ-Energy}$h$) and pressure strain (figure \ref{fig:EQ-Energy}$i$) all show that most of the computed equilibria have much smaller amplitudes than those of the instantaneous chaotic state.} This is consistent with the finding of the DNS study by \cite{doohan_willis_hwang_2021}. However, it should be pointed out that most of their equilibria are featured by the self-sustaining process either at large or small scale, whereas many equilibria in this study have non-negligible amounts of energy and production at both of the scales (figures \ref{fig:EQ-Energy}$a,b,e$) {-- note that the equilibria of \cite{doohan_willis_hwang_2021} were obtained by continuing the existing solutions obtained in a smaller computational domain (equivalent to lower Reynolds number), whereas those in our study are obtained by directly looking for the zeros of (\ref{eq:ROMeq}).} This observation suggests that the relatively low amplitude of the observables associated with energy cascade and nonlinear scale interactions is possibly more associated with the lack of unsteadiness in the equilibria, as we shall indeed see in \S\ref{subsec:PO}.

\subsection{Periodic orbits}\label{subsec:PO}
\begin{figure}
  \centerline{\includegraphics[width = 11cm]{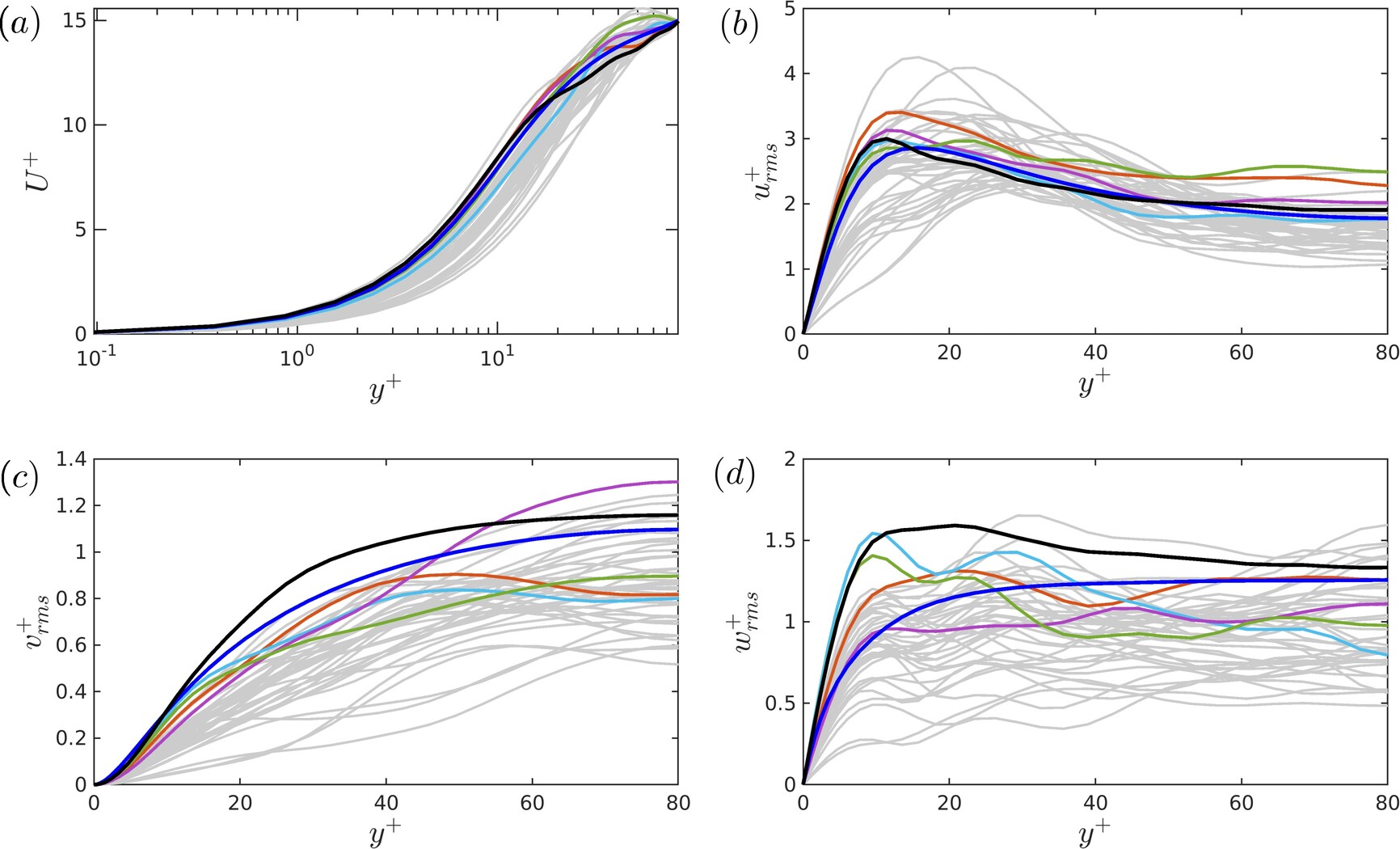}}
  \caption{Mean and root mean square velocity profiles for the periodic orbits (grey) compared to the long-time trajectory generated by the ROM (black) and DNS (blue): $(a)$ $U^+$; $(b)$ $u_{rms}^+$; $(c)$ $v_{rms}^+$; $(d)$ $w_{rms}^+$.  Highlighted periodic orbits: PO17 (cyan); PO20 (purple); PO26 (green); PO42 (orange).}
\label{fig:PO-statistics}
\end{figure}

\begin{figure}
  \centerline{\includegraphics[width = 13cm]{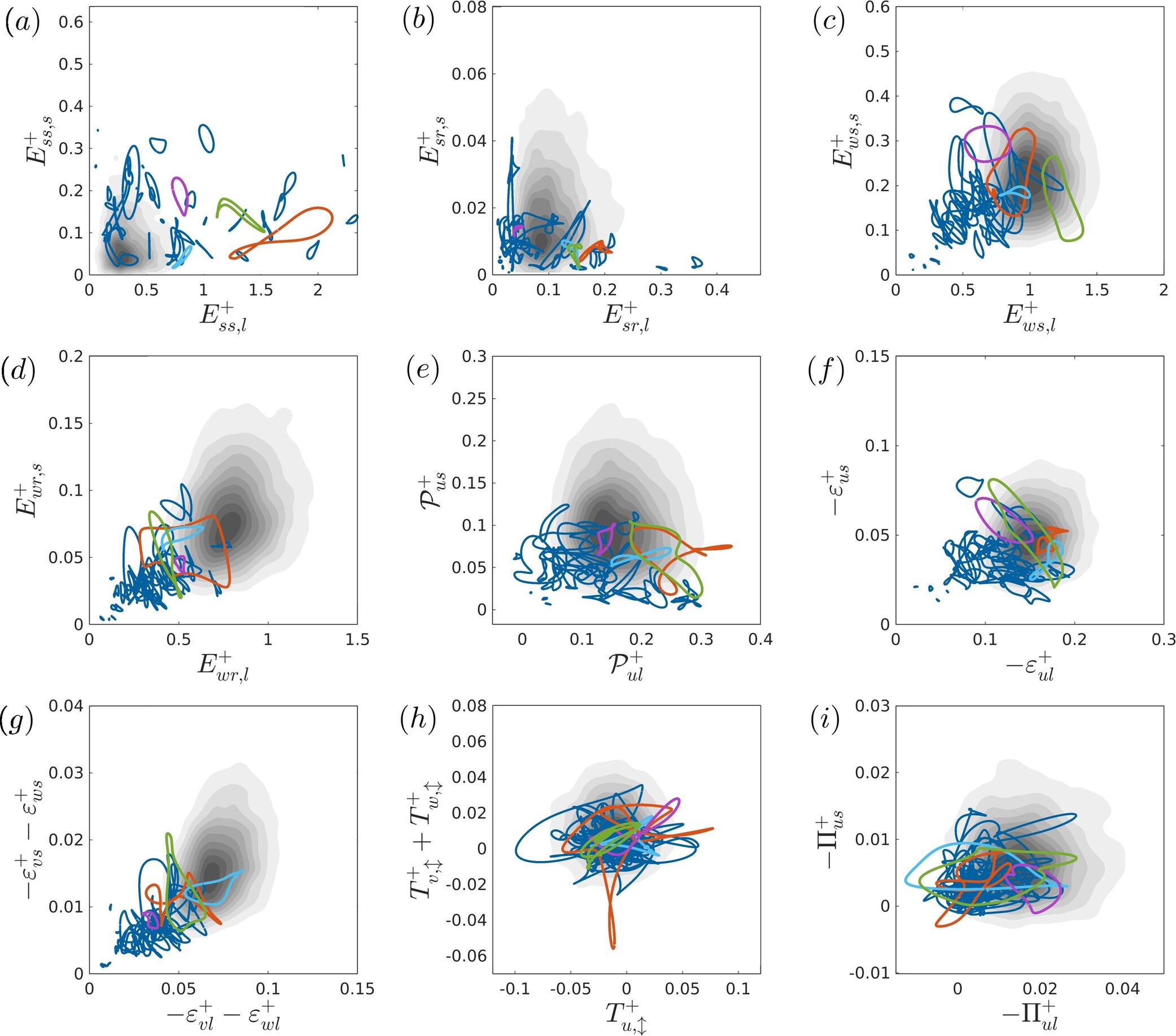}}
  \caption{Phase portraits of the large and small scale energy observables of the self-sustaining process and two-scale energy budget for the periodic orbits (blue), PO17 (cyan), PO20 (purple), PO26 (green), PO42 (orange), {compared to linearly spaced contours of the respective bivariate probability density function (dark-light grey)} generated from a long time trajectory of the ROM.}
\label{fig:PO-Energy}
\end{figure}

Similar data as shown for the equilibria is now shown for the periodic orbits computed using the ROM, with mean and rms velocity profiles shown in figure \ref{fig:PO-statistics}, and phase portraits of large- and small-scale energy observables in figure \ref{fig:PO-Energy}.  Highlighted periodic orbits in figures \ref{fig:PO-statistics} and \ref{fig:PO-Energy} are chosen as they show reasonable agreement to the mean and rms statistics, and their dynamics in the state space will be further explored with respect to figure \ref{fig:PO-Energy}.  

 \begin{figure}
  \centerline{\includegraphics[width = 11cm]{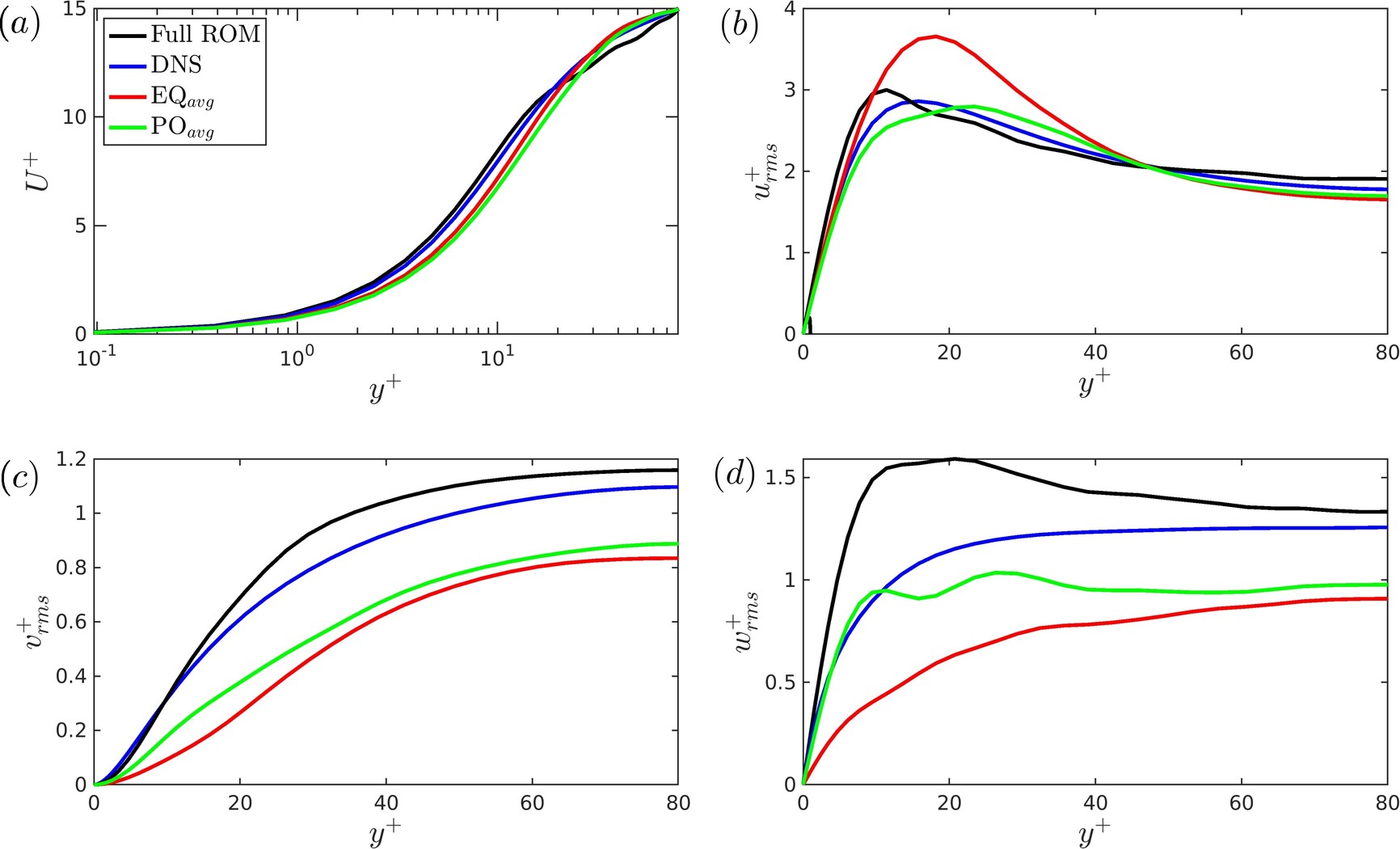}}
  \caption{Mean and root mean square velocity profiles from simple average over the equilibria (red) and the periodic orbits (green) compared to those of the long-time trajectory generated by the ROM (black) and DNS (blue): $(a)$ $U^+$; $(b)$ $u_{rms}^+$; $(c)$ $v_{rms}^+$; $(d)$ $w_{rms}^+$.}
\label{fig:EQ-PO-statistics}
\end{figure}

In general, the periodic orbits characterise the chaotic state more accurately than the equilibria, showing closer agreement to the long time trajectory in many cases. The mean profiles displayed by the periodic orbits appear to be similar to those shown by the equilibria, with the wall shear stress being captured well (figure \ref{fig:PO-statistics}$a$). 
Large maxima seen in the near-wall region of the streamwise rms velocity profiles in the equilibria are less apparent in the periodic orbits with a multitude of periodic orbits capturing this profile well (compare figure \ref{fig:PO-statistics}$b$ with \ref{fig:EQ-statistics}$b$).  Similar to the equilibria, the wall-normal fluctuations displayed by the periodic orbits appear to be mostly smaller than that of the long-time trajectory or DNS, although wall-normal fluctuations in the near-wall region are seen to be a little closer to the long-time trajectory than those of the equilibria (compare figure \ref{fig:PO-statistics}$c$ with \ref{fig:EQ-statistics}$c$).  The spanwise rms profiles also generally exhibit values less than that of the long-time trajectory, although some periodic orbits replicate the near-wall peaks better than those of the equilibria: PO17 and PO26, in particular (compare figure \ref{fig:PO-statistics}$c$ with figure \ref{fig:EQ-statistics}$c$). 

{Some recent studies have suggested that taking simple averages over computed invariant solutions \cite[]{chandler2013invariant,page2022recurrent} is often found to be as good as other more sophisticated approaches \cite[cyclic expansion;][]{Cvitanovic2005}. In figure \ref{fig:EQ-PO-statistics}, turbulence statistics obtained from a simple average of equilibria and periodic orbits are further compared with those of the long time trajectory of DNS and ROM. While there is very little difference in the mean profiles from the simple average of equilibria and periodic orbits (figure \ref{fig:EQ-PO-statistics}$a$), the velocity fluctuations from periodic orbits show better agreements with those from the long time trajectory of DNS and ROM (figures \ref{fig:EQ-PO-statistics}$b$-$d$). However, the velocity fluctuations from periodic orbits still show considerable differences from those from the long time trajectory of ROM, indicating that there are a number of orbits which may not be near the turbulent solution trajectory in the state space.} 

The periodic orbits generally appear to capture the cross-streamwise velocity fluctuations better than the equilibria, as seen in figure \ref{fig:PO-statistics}. The observables associated with energy transport between different velocity components would therefore be better captured by them, and pressure strain has been understood to be mainly involved in this process \cite[e.g.][]{Tennekes1967,cho_hwang_choi_2018}. Given the form of pressure strain shown in (\ref{eq:pres-strain-l}) and (\ref{eq:pres-strain-s}), the state-space projections of the periodic orbits with the observables involving $\partial u/\partial x$ are expected to be improved compared to that of the equilibria. Indeed, an examination of the phase portraits in figure \ref{fig:PO-Energy} reveals that such observables are more convincingly captured in the periodic orbits at both large and small scales than in the equilibria: for example, wavy streaks (figure \ref{fig:PO-Energy}$c$), cross-streamwise dissipation (figure \ref{fig:PO-Energy}$g$) and pressure strain (figure \ref{fig:PO-Energy}$i$), in particular. 

Overall, the periodic orbits describe the chaotic trajectory both statistically and dynamically better than the equilibria. The periodic orbits are seen to be more representative with respect to the energy of straight streaks in the flow, with less high-energy small-scale outliers, although these high-energy outliers continue to be seen in the large scales (figure \ref{fig:PO-Energy}$a$). There is a modest improvement in the description of rolls by the periodic orbits (figures \ref{fig:PO-Energy}$b$,$d$). Production and dissipation appear to be more consistently captured by the periodic orbits than the equilibria (figures \ref{fig:PO-Energy}$e$-$g$). The inter-scale energy transport terms from the periodic obits are also able to explore more of the state space than those from the equilibria. 

Despite the improved state-space dynamics offered by the periodic orbits compared to those by the equilibria, the periodic orbits computed in this study do not sufficiently cover the state space visited by the chaotic trajectory, as is evident from figure \ref{fig:PO-Energy}. {Imposition of the reflection symmetry might be a potential reason, but the phase portrait and turbulence statistics from the chaotic trajectory subject to the symmetry did not yield any significant difference to those without the symmetry (see Appendix \ref{appB}).} It is, however, important to mention that this does not necessarily imply a `theoretical' failure of the attempt to construct the chaotic dynamics in terms of periodic orbits. As mentioned earlier, the periods of the computed orbits are less than $T^+ \leq 141.57$. However, the time scale of the self-sustaining process at large scale is expected to be $T^+ \sim 400-500$ \cite[]{hwang_bengana_2016,doohan_willis_hwang_2021}, considerably longer than the periods of the computed orbits in this study. Therefore, it is presumable that the lack of state-space coverage by the computed orbits originates from their relatively short periods, and the computation of the periodic orbits with longer periods is expected to improve the state-space description by the periodic orbits. Nevertheless, the computation of orbits with longer periods has been found to be challenging even with the present ROM. In particular, we have observed the vanishing recurrence of chaotic trajectory on increasing Reynolds numbers, causing trouble in identifying suitable initial conditions for the Newton algorithm. This also suggests that a numerical approach with a fundamentally different concept would probably be required to overcome this challenge. The Newton method employed for the computation of periodic orbits in this study is expected to have limited applicability, when the leading Lyapunov exponent of the given system is too large -- {the leading Lyapunov exponent of the present ROM computed using the algorithm in \cite{Parker1989} was $\lambda_1=0.107$ ($\lambda_1^+=0.58$), comparable to the leading eigenvalues of the periodic orbits in table \ref{tab:PO-overview}.} This is also a common feature of some of the most popular algorithms currently available based on the same type of Newton iteration \cite[e.g.][]{viswanath2007recurrent,willis2016symmetry}. There have been a few recent works addressing this challenge, and the proposed algorithms do not appear to depend on the magnitude of the leading Lyapunov exponent \cite[e.g.][]{azimi2022constructing,parker2022variational,page2022recurrent}. Employing such a technique may be a way to overcome this challenge in the future. 

\subsection{Multi-scale dynamics in periodic orbits}
Thus far, we have seen that a few periodic orbits computed appear to provide some multi-scale dynamics of the chaotic state, consistent with the results of figure \ref{fig:PO-Energy}, which show contributions in energy terms from both the large and small scales. Now, we will consider some periodic orbits, highlighted with distinct colours in figures \ref{fig:PO-statistics} and \ref{fig:PO-Energy} (PO17, PO20, PO26, PO42, in particular), to explore the exact dynamics that each orbit offers. We shall see that these orbits describe: 1) the self-sustaining process at small and large scales \cite[]{hwang_bengana_2016}; 2) the driving mechanism of small-scale production via energy \cite[]{doohan_willis_hwang_2021}; 3) energy cascade from large to small scales. An overview of these chosen orbits is tabulated in table \ref{tab:PO-overview}. In particular, their stability properties show a large number of unstable modes with a high leading Lyapunov exponent, with values much larger than those reported at low Reynolds numbers \citep{viswanath2007recurrent}.  We also note the large dimension of the unstable manifold of these orbits, comparable to or larger than that of the multi-scale equilibrium solution found in \cite{doohan_2022-tw} (i.e. $L_{L2b}$). Invariant solutions with a high dimensional unstable manifold have often been observed to more closely resemble the turbulent state \cite[see also][]{doohan_willis_hwang_2019}. 

 \begin{table}
  \begin{center}
\def~{\hphantom{0}}
      \begin{tabular}{lcccccc}
         & $\Rey_\tau$ & $T^+$  & $\max\{u_{rms}^+\}$ & $\dim(\mathcal{W}^u)$ & $\mu_1$ & $\lambda_1$ \\[3pt]
    ROM  & 80.2  &  ---   & 3.00 & ---  & --- & {$0.107$} \\
    PO17 & 72.2  & ~61.51 & 2.97  & 97 & -5.4813 & $0.1483$   \\
    PO20 & 78.9  & ~65.11 & 3.13 & 88 & $1.1523 \pm 3.1462\mathrm{i}$ & 0.0995    \\
    PO26 & 75.8  & ~89.26 & 2.96 &  96 & $4.8058 \pm 4.3897\mathrm{i}$ & 0.1125 \\
    PO42 & 78.9  & 124.83 & 3.41 & 103 & $-9.4228 \pm 11.8629\mathrm{i}$ & 0.1167          
    \end{tabular}
  \caption{Friction Reynolds number $\Rey_\tau$ (with respect to the friction velocity of each orbit), period $T^+$ (with respect to the friction velocity of the ROM), maximum streamwise rms velocity $\max\{u_{rms}^+\}$, the dimension of the unstable manifold $\dim(\mathcal{W}^u)$, the largest Floquet multiplier $\mu_1$ and corresponding Lyapunov exponent $\lambda_1$ of selected periodic orbits, compared to the ROM whenever possible.}
  \label{tab:PO-overview}
  \end{center}
\end{table}

\begin{figure}
  \centerline{\includegraphics[width = 12cm]{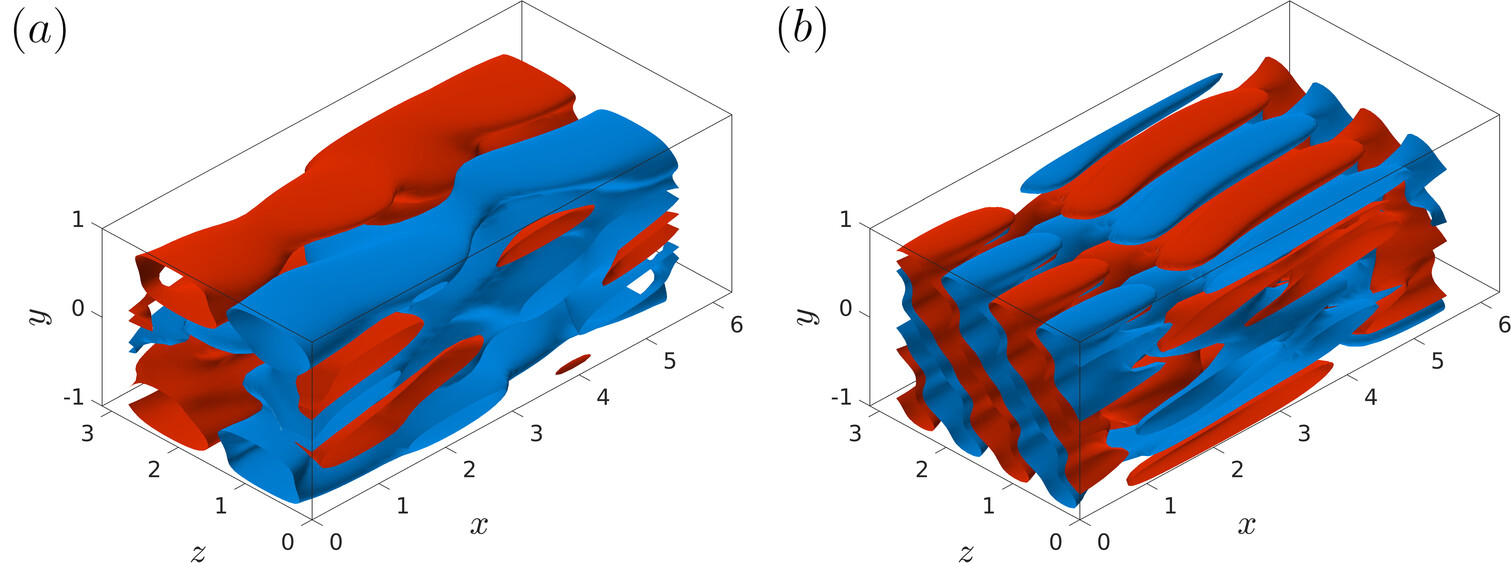}}
  \caption{$(a)$ Instantaneous large scale streamwise velocity isosurfaces $u_l^+= \pm1.5$ (red/blue) and $(b)$ instantaneous small scale streamwise velocity isosurfaces $u_s^+= \pm0.75$ (red/blue) shown at the same instant in time in PO42.}
\label{fig:PO26-large-small}
\end{figure}

Figure \ref{fig:PO26-large-small} shows the large- and small-scale velocity fields at an instant in time in PO42.  It is clear that a large pair of streaks that extend from the wall into the centre of the domain dominate the large-scale field, with twice as many small-scale streaks dominating the near wall of the small-scale velocity field, in agreement with the phenomenology of the large- and small-scale self-sustaining processes. This structure in the large- and small-scale velocity fields, common to a number of the computed periodic orbits, combines to exhibit a rich variety of dynamics.

\begin{figure}
  \centerline{\includegraphics[width = 13cm]{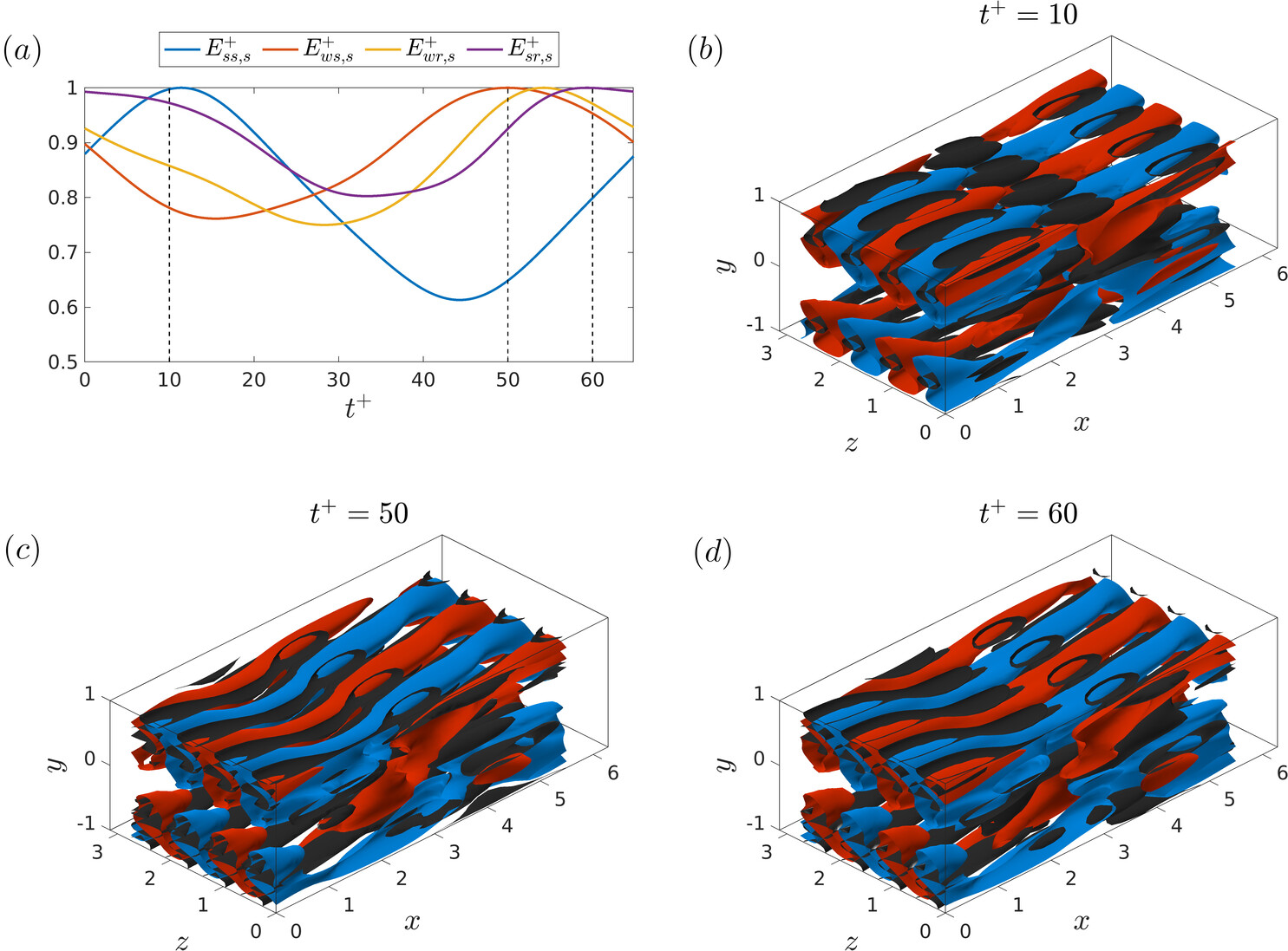}}
  \caption{$(a)$ The evolution of the small-scale self-sustaining process observables normalised by their maximum values over a single period of PO20.  $(b,c,d)$ Flow field visualisations at selected times show small scale streamwise velocity isosurfaces $u_s^+=\pm1.5$ (red/blue), and small scale streamwise vorticity isosurfaces $\omega_{x,s}=\pm0.55$ (black).}
\label{fig:PO20-ssp}
\end{figure}

A number of the spatiotemporal processes statistically identified in the long-time chaotic trajectory have also been found in the periodic orbits. Evidence of the small-scale self-sustaining process has been seen in a large number of the periodic orbits, with an example (PO20) shown in figure \ref{fig:PO20-ssp}.  The time trace of the observables of the small-scale self-sustaining process shows a consistent interchange of dominance between its structural components: $E_{sr,s} \rightarrow E_{ss,s} \rightarrow E_{ws,s} \rightarrow E_{wr,s} \rightarrow E_{sr,s}\,$.  The flow field reconstructions of figure \ref{fig:PO20-ssp}, show the evolution of two pairs of streaks (red/blue $u_s^+$ contours) and vortices (black $\omega_{x,s}$ contours), showing straight streaks undergoing an instability resulting in a wavy streak.  This wavy streak is accompanied by a growth of intensity of the neighbouring streamwise vortices which slowly begin to grow between approximately $t^+=60$ and $t^+=10$ in the absence of streaks.  During this time, straight streaks are seen to regenerate and the cycle continues.

\begin{figure}
  \centerline{\includegraphics[width = 13cm]{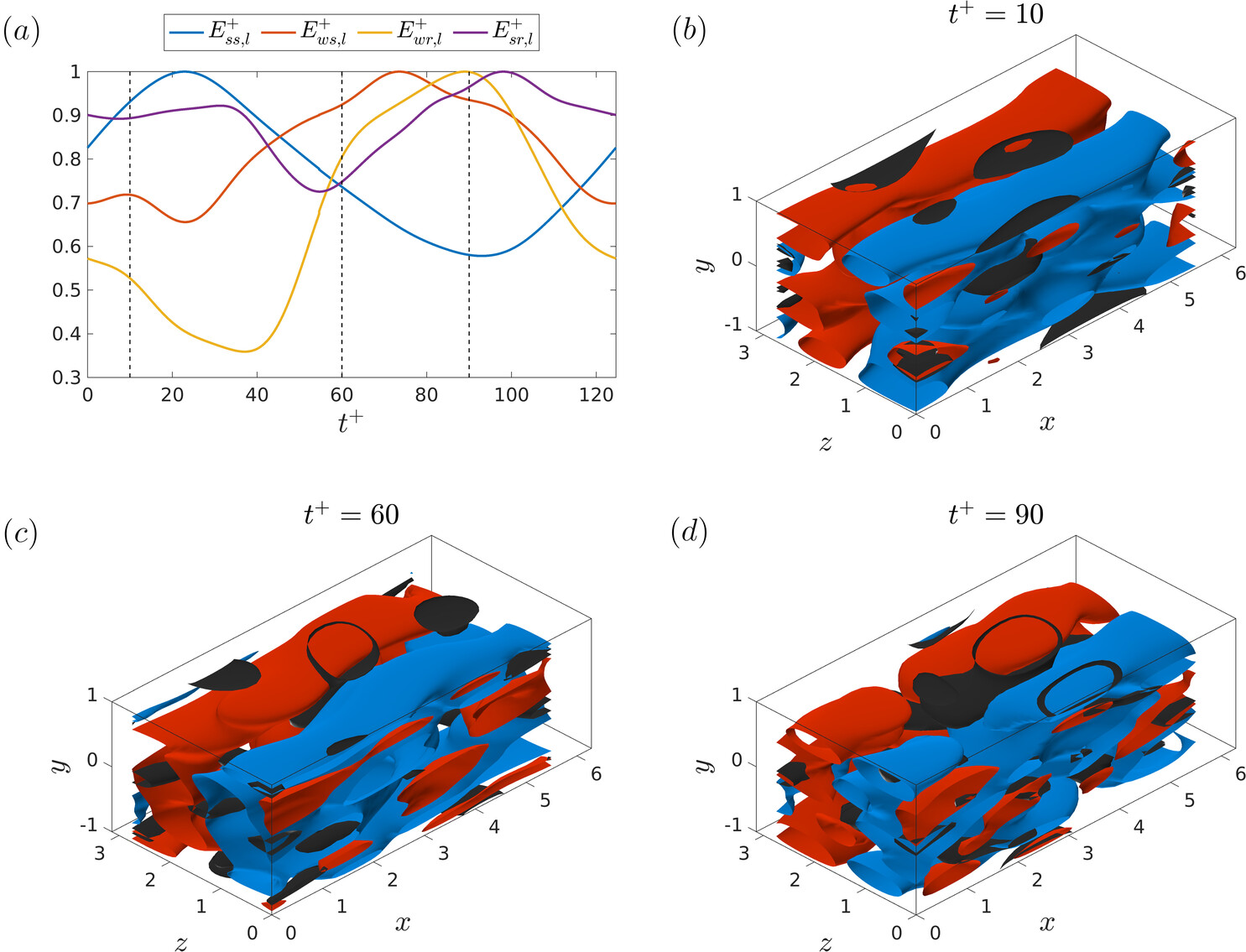}}
  \caption{$(a)$ The evolution of the large-scale self-sustaining process observables normalised by their maximum values over a single period of PO42.  $(b,c,d)$ Flow field visualisations at selected times show large-scale streamwise velocity isosurfaces $u_l^+=\pm1.5$ (red/blue), and large-scale streamwise vorticity isosurfaces $\omega_{x,l}=\pm1$ (black).}
\label{fig:PO42-ssp}
\end{figure}

Due to the increased difficulty and expense of converging longer periodic orbits, evidence of the large-scale self-sustaining process has only been found in a single longer orbit (PO42), shown in figure \ref{fig:PO42-ssp}.  As in the case of the small-scale self-sustaining process, we see a consistent interchange of dominance between the structural components of the large-scale self-sustaining process: $E_{sr,l} \rightarrow E_{ss,l} \rightarrow E_{ws,l} \rightarrow E_{wr,l} \rightarrow E_{sr,l}\,$.  A single pair of large-scale streaks (red/blue $u_l^+$ contours) are seen to dominate the flow field, starting as straight streaks, slowly undergoing an instability to form a wavy streak, resulting in a breakdown in the streak along its length. From this point, a large intensity of large-scale streamwise vorticity is formed which sits alongside the remnants of the broken streak. This vortex is then seen to stretch in the streamwise direction \cite[]{schoppa2002coherent} accompanying the regeneration of the original straight streak occupying the full streamwise length of the domain.  

\begin{figure}
  \centerline{\includegraphics[width = 13cm]{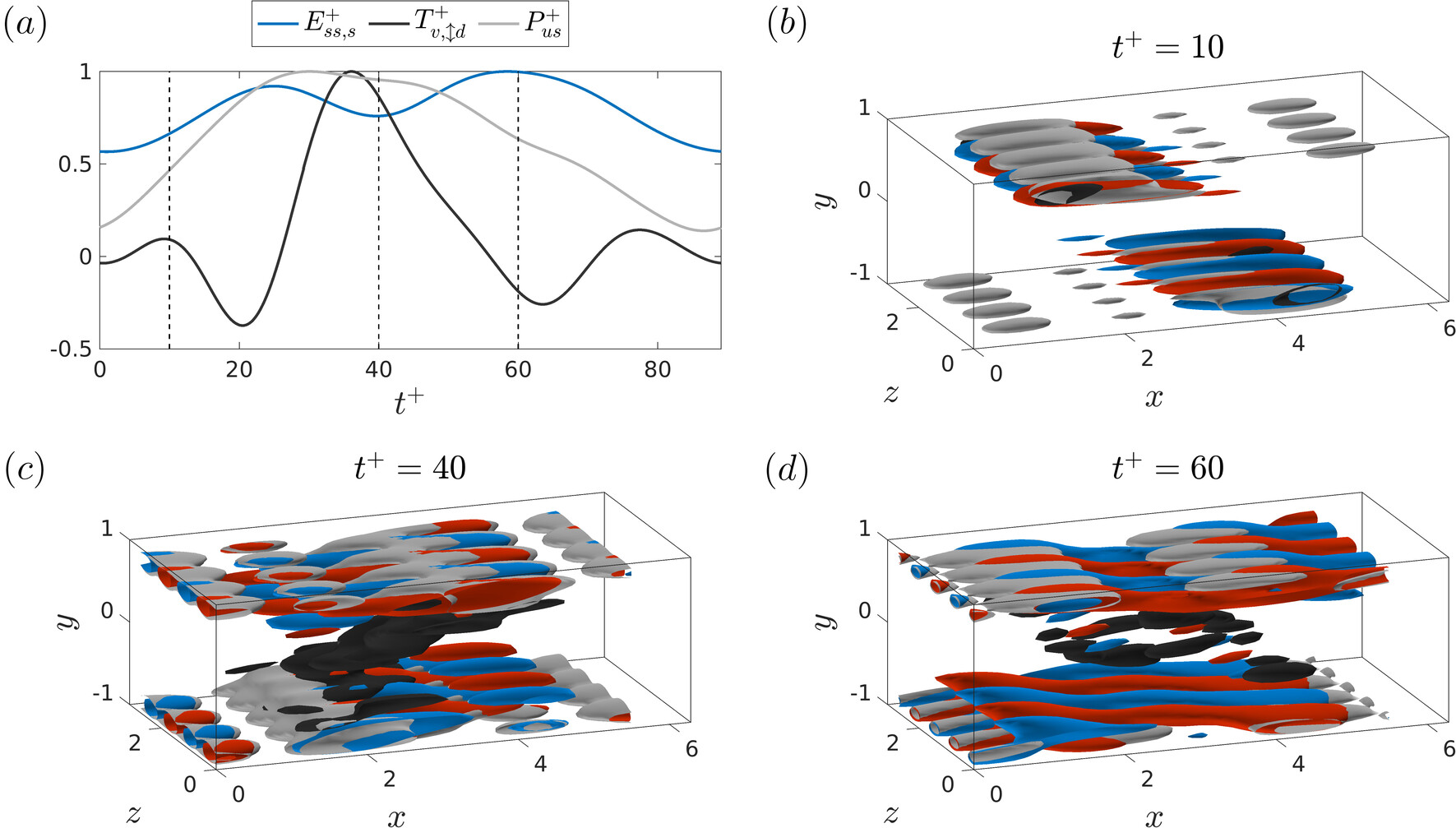}}
  \caption{$(a)$ The evolution of the small scale streak energy $E_{ss,s}^+$, driving component of wall-normal inter-scale turbulent transport $T_{v,\updownarrow d}^+$, and small scale production $P_{us}^+$ normalised by their maximum values over a single period of PO26, showing the driving mechanism. $(b,c,d)$ The flow field visualisations at selected times show small-scale streamwise velocity isosurfaces $u_s^+=\pm2.25$ (red/blue), driving wall-normal inter-scale turbulent transport isosurfaces $T_{v,\updownarrow d}^+=0.11$ (black), and small-scale production isosurfaces $P_{us}^+=0.67 $ (grey).}
\label{fig:PO26-driving}
\end{figure}

Other processes such as the energy cascade and the driving mechanism, identified statistically in long-time chaotic trajectories of the ROM (\S\ref{sec:energy-dynamics}) as well as in the DNS of \citet{doohan_willis_hwang_2021}, have also been identified in a number of the periodic orbits. An example of the driving mechanism is observed in PO26, shown in figure \ref{fig:PO26-driving}.  The first snapshot ($t^+=10$) shows a point of inactivity in the cycle featuring small amounts of small-scale production accompanying some high $u_s^+$ (streaky) structures (figure \ref{fig:PO26-driving}$b$).  By the second snapshot at $t^+=40$, a large peak in the driving wall-normal inter-scale turbulent transport (see (\ref{eq:driving-decomposition})) from large to small scale (i.e. positive $T_{v,\updownarrow d}^+$) is observed both in the centre of the channel and around the streaks, transporting energy from large to small scales. This is accompanied by a rise in turbulent production (figure \ref{fig:PO26-driving}$c$), due to an injection of energy into the wall-normal small-scale velocity $v_s^+$ (see also figure \ref{fig:PO26-driving}$a$). The result of this is an amplification of the small-scale streaky motions observed at $t^+=60$ (figure \ref{fig:PO26-driving}$d$).

\begin{figure}
  \centerline{\includegraphics[width = 13cm]{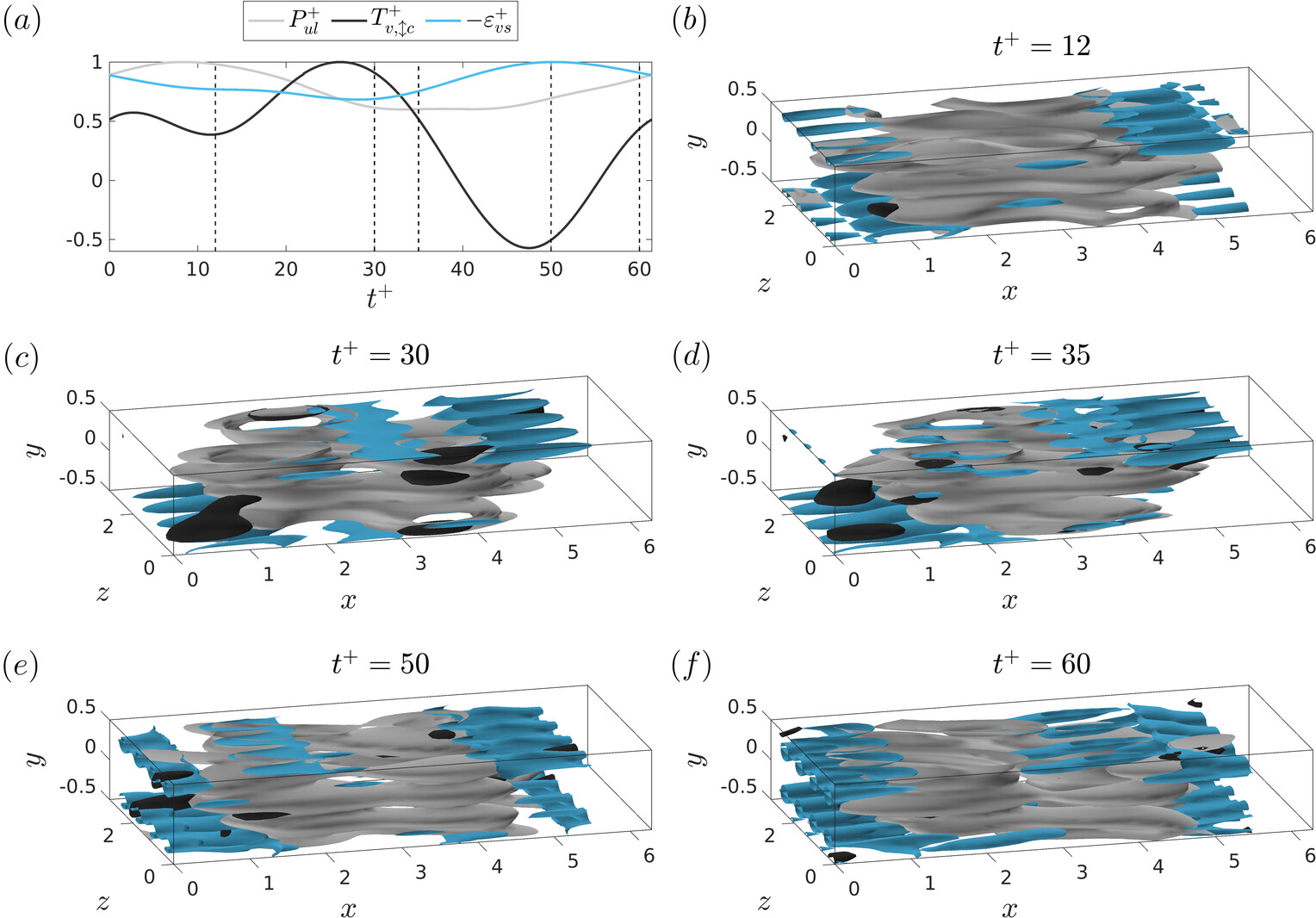}}
  \caption{$(a)$ The evolution of the large-scale production $P_{ul}^+$, cascade component of the wall-normal inter-scale turbulent transport $T_{v,\updownarrow c}^+$, and wall-normal small-scale dissipation $\varepsilon_{vs}^+$ normalised by their maximum values over a single period of PO17, showing the energy cascade.  $(b,c,d)$ The flow field visualisations at selected times show large-scale production isosurfaces $P_{ul}^+=0.22$ (grey), wall-normal inter-scale turbulent transport isosurfaces $T_{v,\updownarrow c}^+=0.22$ (black), and wall-normal and spanwise small-scale dissipation $\varepsilon_{vs}^++\varepsilon_{ws}^+=-0.3$ (cyan). }
\label{fig:PO17-cascade}
\end{figure}

An example of the energy cascade is observed in PO17, shown in figure \ref{fig:PO17-cascade}.  In the statistical analysis of the long-time chaotic trajectory, energy transport primarily through a particular part of the wall-normal inter-scale turbulent transport associated with the cascade $T_{v,\updownarrow c}$ (see (\ref{eq:cascade-decomposition})) was seen to transport energy from the large-scale production to the wall-normal small-scale dissipation near $y \pm 0.5$. Thus, figure \ref{fig:PO17-cascade} shows snapshots of the flow in the central region of the channel. Large-scale production isosurfaces reveal long streak-like structures in the centre of the channel at the peak of the large-scale production at around $t^+=12$ (figure \ref{fig:PO17-cascade}$b$). Between $t^+=12$ and $t^+=30$ these structures shorten slightly in length accompanied by a growth in regions of high wall-normal inter-scale turbulent transport (figure \ref{fig:PO17-cascade}$c$), which are seen to move down and intersect with the small-scale dissipation contours at $t^+=35$, continuing until roughly $t^+=50$ (figures \ref{fig:PO17-cascade}$d$,$e$). At this point, an increase in the wall-normal small-scale dissipation is observed, with a drift of the small-scale dissipation isosurfaces towards the central region of the channel. It is also interesting to note that $T_{v,\updownarrow c}^+<0$ during this time interval, indicating possible existence of an inverse energy cascade, although the time-averaged value of $T_{v,\updownarrow c}^+$ is positive. From $t^+=50$ to $t^+=60$ (figure \ref{fig:PO17-cascade}$f$), the large scale production increases again, correlated to the intensification of large scale straight streaks, completing the cycle.

\section{Conclusions} \label{sec:conclusion}
In this paper, we have validated the model of \citet{Cavalieri-ROM} for plane Couette flow retaining only two integral length scales through a two-part process. The first involved the examination of the time-averaged two-scale energy budget based on a split by spanwise wavenumber, including accessing the exchange of energy within and between scales.  Overall, a qualitative agreement was seen between the ROM and the DNS results of \citet{doohan_willis_hwang_2021} in nearly every respect. The second part of the model validation process involved examining these processes further by studying the temporal cross-correlations of various energy measures integrated over the spatial domain.  Identified in the model, was both a large- and small-scale self-sustaining process, an energy cascade from large to small scales, the driving of small-scale production via energy transport from the large scales in the near-wall region, and a feeding process transporting energy from the small to large scales. The underlying mechanisms of these processes, at least in relation to the various energy budget terms and the structural components of the self-sustaining process at each scale, have been analysed and compared to the results of \citet{doohan_willis_hwang_2021}, with good qualitative agreement. 

96 equilibria and 43 periodic orbits have subsequently been computed for the ROM, the statistics and phase portraits of which have been directly compared to those of a long-time chaotic trajectory generated using the ROM. In general, the invariant solutions of the ROM reproduce its multi-scale dynamics more convincingly than those found by \citet{doohan_2022-tw} for the full Navier-Stokes equations. Equilibrium solutions were largely seen to capture straight streaks and rolls, resulting in large near-wall peaks in the streamwise rms velocity profiles, with high $\Rey_\tau$ equilibria seen to exhibit energy production values very similar to that of the mean long-time trajectory.  In general, equilibria failed to accurately capture wavy streaks and rolls, as well as wall-normal and spanwise energy dissipation. The periodic orbits were seen to characterise the chaotic state more accurately than the equilibria, showing closer agreement to the long-time trajectory in nearly all cases, exhibiting a rich range of multi-scale dynamics and spatiotemporal evolution. 

The periodic orbits have particularly been seen to contain useful dynamical information about the known processes in the flow, which have been more easily identified than by using a statistical approach applied to the long-time chaotic trajectory. In particular, various periodic orbits have contained valuable information, including the large- and small-scale self-sustaining process, the driving mechanism which transports energy from the large scales to the small-scale production, and the energy cascade, allowing for a more clear physical interpretation of these processes. {It might be seen evident that periodic orbits would be more useful and relevant than equilibrium solutions for the statistical and dynamical description of turbulence. However, it is probably worth pointing out that, at lower Reynolds numbers, where the flow is characterised by a single integral length scale (i.e. the minimal flow unit), many equilibrium solutions in plane Couette flow have also previously been found to be organised around the turbulent trajectory in the state space \cite[e.g.][]{gibson2008visualizing,doohan_willis_hwang_2019}. These equilibrium solutions often contain the most fundamental physical ingredients for the underlying physical mechanism of sustaining turbulence (i.e. the self-sustaining process). Similarly, in Rayleigh-B\'enard convection, the recent study by \cite{motoki2021multi} showed that there exists a multi-scale equilibrium solution which resembles turbulence. Even from a dynamical systems viewpoint, the equilibrium solutions can still act as a signpost for the chaotic solution trajectory through the formation of homoclinic/heteroclinic orbits. From this perspective, it is interesting to observe that a large number of the equilibrium solutions in a wall-bounded shear flow lose their statistical and dynamical importance on increasing the Reynolds number. It remains to be seen if this is a general feature in other flows.}

As a final remark, the invariant solutions computed using the ROM are not invariant solutions of the exact system, and this is an important limitation of this study. However, the computation of invariant solutions to the full Navier-Stokes equations is prohibitively expensive and very challenging for high Reynolds numbers. The present study has demonstrated that the significantly reduced cost of generating invariant solutions for a carefully crafted ROM would provide a gateway to apply dynamical systems approaches to flows at high Reynolds numbers and explore increasingly complex multi-scale turbulent dynamics identified the full-scale dynamics {-- indeed, the dramatically reduced computational cost has enabled us to carry out an extensive search of unstable periodic orbits buried in turbulent state, a task that is likely impossible to tackle using DNS at the current computational cost for the given Reynolds number.} {In this respect, the key to the application of this approach is in the construction of a successful ROM, which is able to capture the turbulent dynamics of interest. Thankfully, there are an increasing number of studies towards the reliable construction of ROMs for turbulent flows \cite[e.g. see the review by][]{Rowley2017}. For the purpose of extracting the key turbulent dynamics through invariant solutions, the ROM must be in the form of a `nonlinear' dynamical system, and it is highly desirable to preserve the mathematical structure of the Navier-Stokes equations so that the dynamics described by invariant solutions of the ROM can be analysed in comparison to that of the real flow. By doing so, the invariant solutions would be able to describe the exact `reduced-order' dynamics of the given turbulent flow and the underlying complex physical processes can be studied by formulating a suitable energy budget analysis, a key tool to explore energy cascade and scale interactions in turbulence. From this perspective, it is also important to carefully control the construction of ROM by recognising its exact capability in the description of turbulent dynamics.} Lastly, it is worth noting that the invariant solutions from a ROM could be continued with respect to the neglected modes of the exact system, provided that the ROM has a sufficient number of degrees of freedom. This is a direction we currently pursue towards a cost-effective method for the computation of physically meaningful invariant solutions to the Navier-Stokes equations.

\backsection[Acknowledgements]{This paper was initiated while A.V.G.C. and Y.H. were visiting the Isaac Newton Institute for Mathematical Sciences at the University of Cambridge for the
programme ‘Mathematical aspects of turbulence: where do we stand?’. We would like to thank the organisers of the programme.}

\backsection[Funding]{This work was supported by the financial support of the Engineering and Physical Sciences Research Council (EPSRC; EP/T009365/1) in the UK. Visit of A.V.G.C. and Y.H. to the Isaac Newton Institute for Mathematical Sciences at the University of Cambridge was supported by EPSRC through the programme ‘Mathematical aspects of turbulence: where do we stand?’ (EP/R014604/1). M.M. is supported by the Priority Programme SPP 1881 ``Turbulent Superstructures" of the Deutsche Forschungsgemeinschaft (DFG, grant number Li3694/1) and by the School of Mathematics at the University of Edinburgh.}

\backsection[Declaration of interests]{The authors report no conflict of interest.}

\appendix

\section{Terms of the Two-Scale Energy Budget}\label{appA}
 In \S \ref{sec:Energy_Budget_Chp}, the terms of the two-scale energy budget i.e. equation \ref{eq:energy-eq-large}, \ref{eq:energy-eq-small} were introduced. The precise terms in these equations are summarised here. 

The large-scale kinetic energies are given by,
\begin{equation}
    E_{ul} = \tfrac{1}{2}\big\langle u_{l}^2 \big\rangle_{x,z} \quad , \quad E_{vl} = \tfrac{1}{2}\big\langle v_{l}^2 \big\rangle_{x,z} \quad, \quad E_{wl} = \tfrac{1}{2}\big\langle w_{l}^2 \big\rangle_{x,z},
\end{equation}
the large-scale production is,
\begin{equation}
    P_{ul} = -\frac{\partial U}{\partial y} \langle u_l\, v_l \rangle_{x,z},
\end{equation}
the large-scale turbulent transport terms are,
\begin{subequations}
\begin{alignat}{4}\label{eq:transport-l}
& T_{ul} = -\langle  u_l (\boldsymbol{u}_l\bcdot \nabla u_l &&+ \boldsymbol{u}_l\bcdot \nabla u_s &&+ \boldsymbol{u}_s\bcdot \nabla u_l &&+ \boldsymbol{u}_s\bcdot \nabla u_s) \rangle_{x,z},  \\
& T_{vl} = -\langle  v_l (\boldsymbol{u}_l\bcdot \nabla v_l &&+ \boldsymbol{u}_l\bcdot \nabla v_s &&+ \boldsymbol{u}_s\bcdot \nabla v_l &&+ \boldsymbol{u}_s\bcdot \nabla v_s) \rangle_{x,z},  \\
& T_{wl} = -\langle  w_l (\boldsymbol{u}_l\bcdot \nabla w_l &&+ \boldsymbol{u}_l\bcdot \nabla w_s &&+ \boldsymbol{u}_s\bcdot \nabla w_l &&+ \boldsymbol{u}_s\bcdot \nabla w_s) \rangle_{x,z},
\end{alignat}
\end{subequations}
the large-scale pressure strain terms are,
\begin{equation}
    \Pi_{ul} = \Bigl \langle p_l \frac{\partial u_l}{\partial x} \Bigr \rangle_{x,z} \quad,\quad \Pi_{vl} = \Bigl \langle p_l \frac{\partial v_l}{\partial y} \Bigr \rangle_{x,z} \quad,\quad \Pi_{wl} = \Bigl \langle p_l \frac{\partial w_l}{\partial z} \Bigr \rangle_{x,z}, \:
    \label{eq:pres-strain-l}
\end{equation}
the large-scale viscous transport terms are,
\begin{equation}
    T_{\nu,ul} = \frac{1}{2} \frac{1}{\Rey}\Bigl \langle \frac{\partial^2}{\partial y^2}\left( u_l^2 \right) \Bigr \rangle_{x,z} \; \;,\; \; T_{\nu,vl} = \frac{1}{2} \frac{1}{\Rey}\Bigl \langle \frac{\partial^2}{\partial y^2}\left( v_l^2 \right) \Bigr \rangle_{x,z} \; \;,\; \; T_{\nu,wl} = \frac{1}{2} \frac{1}{\Rey}\Bigl \langle \frac{\partial^2}{\partial y^2}\left( w_l^2 \right) \Bigr \rangle_{x,z},
\end{equation}
the large-scale pressure transport is,
\begin{equation}
    T_{p,vl} = -\Bigl \langle \frac{\partial}{\partial y}\left( p_l \, v_l \right) \Bigr \rangle_{x,z} ,
    \label{pres-trans-l}
\end{equation}
and the large-scale dissipation terms are,
\vspace{2pt}
\begin{equation}
    \varepsilon_{ul} = -\frac{1}{Re} \big\langle \nabla u_l \bcdot \nabla u_l \big\rangle_{x,z} \quad,\quad \varepsilon_{vl} = -\frac{1}{\Rey} \big\langle \nabla v_l \bcdot \nabla v_l \big\rangle_{x,z} \quad,\quad \varepsilon_{wl} = -\frac{1}{\Rey} \big\langle \nabla w_l \bcdot \nabla w_l \big\rangle_{x,z}. \:
\end{equation}

Similarly, the small-scale kinetic energies are given by,
\begin{equation}
    E_{us} = \tfrac{1}{2}\big\langle u_{s}^2 \big\rangle_{x,z} \quad , \quad E_{vs} = \tfrac{1}{2}\big\langle v_{s}^2 \big\rangle_{x,z} \quad, \quad E_{ws} = \tfrac{1}{2}\big\langle w_{s}^2 \big\rangle_{x,z},
\end{equation}
\vspace{-10pt}
the small-scale production is,
\vspace{10pt}
\begin{equation}
    {P}_{us} = -\frac{\partial U}{\partial y} \langle u_s\, v_s \rangle_{x,z},
\end{equation}
\vspace{-5pt}
the small-scale turbulent transport terms are,
\vspace{5pt}
\begin{subequations}
\begin{alignat}{4}\label{eq:transport-s}
& T_{us} = -\langle  u_s (\boldsymbol{u}_l\bcdot \nabla u_l &&+ \boldsymbol{u}_l\bcdot \nabla u_s &&+ \boldsymbol{u}_s\bcdot \nabla u_l &&+ \boldsymbol{u}_s\bcdot \nabla u_s) \rangle_{x,z},  \\
& T_{vs} = -\langle  v_s (\boldsymbol{u}_l\bcdot \nabla v_l &&+ \boldsymbol{u}_l\bcdot \nabla v_s &&+ \boldsymbol{u}_s\bcdot \nabla v_l &&+ \boldsymbol{u}_s\bcdot \nabla v_s) \rangle_{x,z},  \\
& T_{ws} = -\langle  w_s (\boldsymbol{u}_l\bcdot \nabla w_l &&+ \boldsymbol{u}_l\bcdot \nabla w_s &&+ \boldsymbol{u}_s\bcdot \nabla w_l &&+ \boldsymbol{u}_s\bcdot \nabla w_s) \rangle_{x,z}, 
\end{alignat}
\end{subequations}
the small-scale pressure strain terms are,
\begin{equation}
    \Pi_{us} = \Bigl \langle p_s \frac{\partial u_s}{\partial x} \Bigr \rangle_{x,z} \quad,\quad \Pi_{vs} = \Bigl \langle p_s \frac{\partial v_s}{\partial y} \Bigr \rangle_{x,z} \quad,\quad \Pi_{ws} = \Bigl \langle p_s \frac{\partial w_s}{\partial z} \Bigr \rangle_{x,z}, \:
    \label{eq:pres-strain-s}
\end{equation}
the small-scale viscous transport terms are,
\begin{equation}
    T_{\nu,us} = \frac{1}{2} \frac{1}{\Rey}\Bigl \langle \frac{\partial^2}{\partial y^2}\left( u_s^2 \right) \Bigr \rangle_{x,z} \; \;,\; \; T_{\nu,vs} = \frac{1}{2} \frac{1}{\Rey}\Bigl \langle \frac{\partial^2}{\partial y^2}\left( v_s^2 \right) \Bigr \rangle_{x,z} \; \;,\; \; T_{\nu,ws} = \frac{1}{2} \frac{1}{\Rey}\Bigl \langle \frac{\partial^2}{\partial y^2}\left( w_s^2 \right) \Bigr \rangle_{x,z},
\end{equation}
the small-scale pressure transport is,
\begin{equation}
    T_{p,vs} = -\Bigl \langle \frac{\partial}{\partial y}\left( p_s \, v_s \right) \Bigr \rangle_{x,z} ,
    \label{pres-trans-s}
\end{equation}
and the small-scale dissipation terms are,
\vspace{2pt}
\begin{equation}
    \varepsilon_{us} = -\frac{1}{\Rey} \big\langle \nabla u_s \bcdot \nabla u_s \big\rangle_{x,z} \quad,\quad \varepsilon_{vs} = -\frac{1}{\Rey} \big\langle \nabla v_s \bcdot \nabla v_s \big\rangle_{x,z} \quad,\quad \varepsilon_{ws} = -\frac{1}{\Rey} \big\langle \nabla w_s \bcdot \nabla w_s \big\rangle_{x,z}. \:
\end{equation}

In addition to this, the turbulent transport terms were further decomposed in equations \ref{eq:transport-decomposition}.  The intra-scale spatial turbulent transport terms are given by,
\begin{subequations}
\begin{alignat}{2}
& T_{ul,-} = -\langle  u_l (\boldsymbol{u}_l\bcdot \nabla u_l) \rangle_{x,z}  &&= -\nabla\bcdot\langle  \tfrac{1}{2}u_l^2\,\boldsymbol{u}_l \rangle_{x,z}, \\
& T_{vl,-} = -\langle  v_l (\boldsymbol{u}_l\bcdot \nabla v_l) \rangle_{x,z}  &&= -\nabla\bcdot\langle  \tfrac{1}{2}v_l^2\,\boldsymbol{u}_l \rangle_{x,z}, \\
& T_{wl,-} = -\langle  w_l (\boldsymbol{u}_l\bcdot \nabla w_l) \rangle_{x,z}  &&= -\nabla\bcdot\langle  \tfrac{1}{2}w_l^2\,\boldsymbol{u}_l \rangle_{x,z}, \\
& T_{us,-} = -\langle  u_s (\boldsymbol{u}_s\bcdot \nabla u_s) \rangle_{x,z}  &&= -\nabla\bcdot\langle  \tfrac{1}{2}u_s^2\,\boldsymbol{u}_s \rangle_{x,z}, \\
& T_{vs,-} = -\langle  v_s (\boldsymbol{u}_s\bcdot \nabla v_s) \rangle_{x,z}  &&= -\nabla\bcdot\langle  \tfrac{1}{2}v_s^2\,\boldsymbol{u}_s \rangle_{x,z}, \\
& T_{ws,-} = -\langle  w_s (\boldsymbol{u}_s\bcdot \nabla w_s) \rangle_{x,z}  &&= -\nabla\bcdot\langle  \tfrac{1}{2}w_s^2\,\boldsymbol{u}_s \rangle_{x,z}.
\end{alignat}
\end{subequations}
The inter-scale spatial turbulent transport terms are given by,
\begin{subequations}
\begin{alignat}{2}
& T_{ul,\#} = -\nabla\bcdot\langle  \tfrac{1}{2}u_l^2\,\boldsymbol{u}_s \rangle_{x,z} &&-\nabla\bcdot\langle  u_l\,u_s\,\boldsymbol{u}_s \rangle_{x,z},\\
& T_{vl,\#} = -\nabla\bcdot\langle  \tfrac{1}{2}v_l^2\,\boldsymbol{u}_s \rangle_{x,z} &&-\nabla\bcdot\langle  v_l\,v_s\,\boldsymbol{u}_s \rangle_{x,z},\\
& T_{wl,\#} = -\nabla\bcdot\langle  \tfrac{1}{2}w_l^2\,\boldsymbol{u}_s \rangle_{x,z} &&-\nabla\bcdot\langle  w_l\,w_s\,\boldsymbol{u}_s \rangle_{x,z},\\
& T_{us,\#} = -\nabla\bcdot\langle  \tfrac{1}{2}u_s^2\,\boldsymbol{u}_l \rangle_{x,z} &&-\nabla\bcdot\langle  u_l\,u_s\,\boldsymbol{u}_l \rangle_{x,z},\\
& T_{vs,\#} = -\nabla\bcdot\langle  \tfrac{1}{2}v_s^2\,\boldsymbol{u}_l \rangle_{x,z} &&-\nabla\bcdot\langle  v_l\,v_s\,\boldsymbol{u}_l \rangle_{x,z},\\
& T_{ws,\#} = -\nabla\bcdot\langle  \tfrac{1}{2}w_s^2\,\boldsymbol{u}_l \rangle_{x,z} &&-\nabla\bcdot\langle  w_l\,w_s\,\boldsymbol{u}_l \rangle_{x,z}.
\end{alignat}
\end{subequations}
The inter-scale turbulent transport terms are given by,
\begin{subequations}
\label{eq:T-arrow}
\begin{alignat}{2}
& T_{u,\updownarrow} = \langle  u_l(\boldsymbol{u}_l\bcdot \nabla u_s) \rangle_{x,z} &&-\langle  u_s(\boldsymbol{u}_s\bcdot \nabla u_l) \rangle_{x,z},\\
& T_{v,\updownarrow} = \langle  v_l(\boldsymbol{u}_l\bcdot \nabla v_s) \rangle_{x,z} &&-\langle  v_s(\boldsymbol{u}_s\bcdot \nabla v_l) \rangle_{x,z},\\
& T_{w,\updownarrow} = \langle  w_l(\boldsymbol{u}_l\bcdot \nabla w_s) \rangle_{x,z} &&-\langle  w_s(\boldsymbol{u}_s\bcdot \nabla w_l) \rangle_{x,z}.
\end{alignat}
\end{subequations}

 \begin{figure}
  \centerline{\includegraphics[width = 11cm]{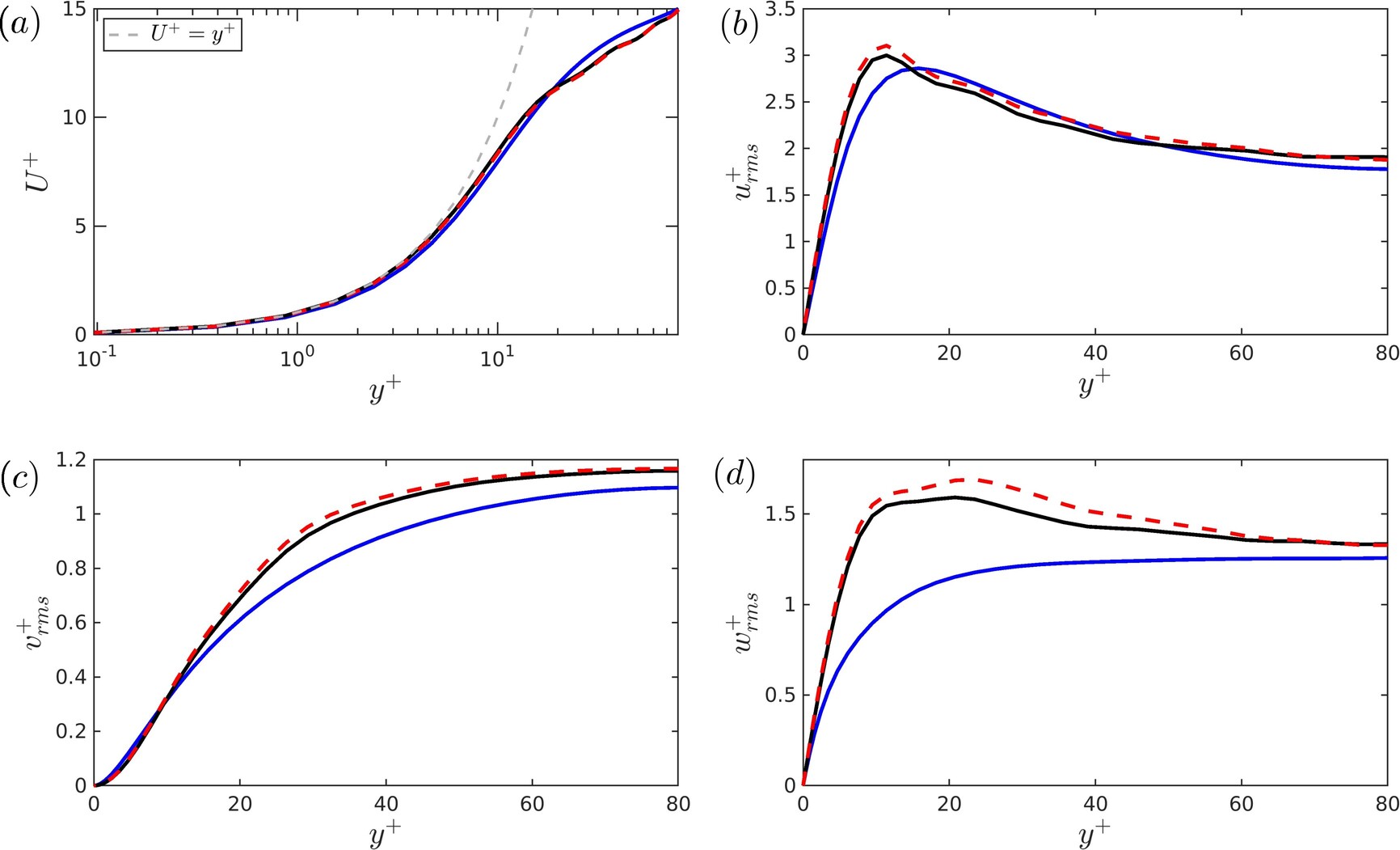}}
  \caption{Mean and root mean square velocity profiles of the long-time trajectory: $(a)$ $U^+$; $(b)$ $u_{rms}^+$; $(c)$ $v_{rms}^+$; $(d)$ $w_{rms}^+$.  Here: DNS (blue); full ROM (black); ROM with the reflection symmetry (red dashed).}
\label{fig:DNS-statistics-sym}
\end{figure}

{\section{Effect of reflection symmetry}\label{appB}
For the search of invariant solutions, the reflection symmetry was imposed for the ROM as discussed in \S\ref{sec:EQ-PO}. Here, its effect on turbulence statistics is reported. Figure \ref{fig:DNS-statistics-sym} shows turbulence statistics from the full ROM and the ROM with the reflection symmetry. All the one-point turbulence statistics (i.e. mean and velocity fluctuations) from the ROM without and with the reflection symmetry are almost identical, indicating that the imposed symmetry is not significantly restrictive. The same behaviour has also been found in the various phase portraits examined in this study (not shown).} 

\bibliographystyle{jfm}
\bibliography{jfm,references}

\end{document}